\documentclass[twocolumn,showpacs,superscriptaddress,preprintnumbers,amsmath,amssymb]{revtex4}

\usepackage{graphicx}
\usepackage{dcolumn}
\usepackage{bm}

\usepackage{amssymb}
\usepackage{latexsym}

\newcommand{\ee}{\end{equation}} 
\newcommand{\be}{\begin{equation}} 
\def\spose#1{\hbox to 0pt{#1\hss}}
\def\ltapprox{\mathrel{\spose{\lower 3pt\hbox{$\mathchar"218$}}
 \raise 2.0pt\hbox{$\mathchar"13C$}}}
\def\gtapprox{\mathrel{\spose{\lower 3pt\hbox{$\mathchar"218$}}
 \raise 2.0pt\hbox{$\mathchar"13E$}}}

\begin{document}


\title{Landau-gauge propagators in Yang-Mills theories at $\beta = 0$: \\[2mm]
       massive solution versus conformal scaling}

\author{A. Cucchieri}
\affiliation{Instituto de F\'{\i}sica de S\~ao Carlos, Universidade de
             S\~ao Paulo, \\ Caixa Postal 369, 13560-970 S\~ao Carlos, SP, Brazil \\[0.5mm]}
\author{T. Mendes}
\affiliation{Instituto de F\'{\i}sica de S\~ao Carlos, Universidade de
             S\~ao Paulo, \\ Caixa Postal 369, 13560-970 S\~ao Carlos, SP, Brazil \\[0.5mm]}
\affiliation{DESY--Zeuthen, Platanenallee 6, 15738 Zeuthen, Germany}

\date{\today}

\begin{abstract}
We study Landau-gauge gluon and ghost propagators in Yang-Mills theories at lattice
parameter $\beta = 0$, considering relatively large lattice volumes for the case
of the SU(2) 
gauge group in three and four space-time dimensions. We compare the lattice data 
to the so-called massive and conformal-scaling solutions, examining the requirements
for a good description of the propagators over various ranges of momenta and
discussing possible systematic errors.
Our analysis strongly supports the massive solution, i.e.\ a finite gluon propagator 
and an essentially free ghost propagator in the infrared limit, in disagreement
with Ref.\ \cite{Sternbeck:2008mv}. Moreover, we argue that discretization effects
play no role in the analysis of these propagators.
\end{abstract}

\maketitle


\section{Introduction}
\label{sec:intro}

In recent years, considerable effort has been invested in the study of
the infrared (IR) behavior of Green's functions in Yang-Mills
theories. The results of these studies, using analytic methods
as well as numerical lattice simulations, are usually compared
to the predictions of various confinement scenarios. Since Green's
functions are gauge-dependent quantities, one can expect to find
different confinement pictures when considering different choices
of the gauge.

Here we consider the Landau gauge and test the predictions of the 
Gribov-Zwanziger and Kugo-Ojima confinement scenarios at lattice parameter
$\beta = 0$, i.e.\ in the limit of infinite lattice coupling. This
study is very similar to that presented in Refs.\ \cite{Sternbeck:2008mv,
Sternbeck:2008wg,Sternbeck:2008na}. On the other hand, it is not
intended to be a simple duplication of that work since, in particular,
the analysis we present here is not done exactly in the same way
as in Ref.\ \cite{Sternbeck:2008mv}. Also, even though there is probably 
no real difference at the level of the lattice data, as we show in 
Section \ref{sec:plots} and explain in Sections \ref{sec:recent0} and
\ref{sec:conclusions}, we mostly do not agree with the data analysis and 
the interpretation of the results presented in Ref.\ \cite{Sternbeck:2008mv}. 
The interested reader is
of course invited to read both papers and form her/his own opinion
on this subject.


\subsection{Confinement scenarios in Landau gauge}

In Landau gauge, the Gribov-Zwanziger scenario \cite{Gribov:1977wm,
Zwanziger:1993dh,Zwanziger:2009je,Cucchieri:2008yp} relates confinement of
quarks to a ghost propagator $G(p)$ enhanced in the IR limit when compared
to the tree-level behavior $1/p^2$ as a function of the momentum $p$. 
Indeed, in this scenario, the
enhancement of $G(p)$ should account for the long-range mechanism
responsible for confinement. An enhanced ghost propagator is also
obtained by the Kugo-Ojima color-confinement criterion \cite{Kugo:1979gm,
Kugo:1995km}. Consequently, in both cases 
one expects to find $\lim_{p\to0} p^2 G(p) = \infty$,
which is referred to as {\em ghost dominance}
\cite{Zwanziger:2002ia, Alkofer:2000wg}.

At the same time, the gluons should be confined due to the violation
of reflection positivity \cite{Zwanziger:1990by}. This implies that the
gluon propagator in position space $D(x)$ should be negative for a range
of values of the space-time separation $x$. Since the gluon propagator at
zero momentum $D(p=0)$ is proportional to $\int d^dx \, D(x)$, it is
clear that negative values for $D(x)$ will tend to reduce the value of
$D(p=0)$, leading to a suppression of $D(p)$ at small momenta
\cite{Gribov:1977wm,Zwanziger:1990by,Dell'Antonio:1989jn}. Similarly,
in the Kugo-Ojima scenario one can show that the perturbative massless
pole of the gluon propagator probably disappears as a consequence of the
confining criterion \cite{Kugo:1995km}. Thus, an IR-suppressed gluon
propagator can also be accommodated in the Kugo-Ojima confinement
scenario \cite{Alkofer:2000wg}. Clearly, maximal violation of reflection
positivity is obtained if $D(p=0)=0$.

Even though these two scenarios predict similar behaviors for the
gluon and ghost propagators in Landau gauge, one should recall that
the line of thinking in the two cases is quite different. Indeed,
in the Gribov-Zwanziger scenario confinement is due to the properties
of the configurations belonging to the boundary of the so-called 
first Gribov region \cite{Gribov:1977wm,Zwanziger:1993dh,Cucchieri:2008yp},
which should be the relevant configurations in Landau gauge.
On the contrary, in the Kugo-Ojima scenario \cite{Kugo:1979gm,Kugo:1995km},
confinement is obtained if one can define unbroken global color charges.
Moreover, in the latter case, Gribov copies do not seem to play any role
in the confining mechanism. Nevertheless, if one uses the Kugo-Ojima
confinement criterion as a boundary condition \cite{Dudal:2009xh} then the
partition function is equivalent to the one obtained using the Gribov-Zwanziger
approach, i.e.\ by restricting the functional integration to
the first Gribov region. 


\subsection{Massive and conformal-scaling solutions}

Yang-Mills theories can be studied non-perturbatively using
the field equations of motion of the theory, i.e. the sets of
coupled Dyson-Schwinger equations (DSE) \cite{Roberts:1994dr}.
Usually, the solution of these equations depends on the considered
truncation scheme and on the approximations employed.

Recently, it has been shown \cite{Alkofer:2008jy,Fischer:2006vf}
that there exist two possible consistent solutions of the DSE
in 4d Landau gauge. One solution,
usually called {\em conformal scaling}, gives an IR-enhanced
ghost propagator $G(p) \sim p^{-2(1+\kappa_G)}$ and an IR-finite
gluon propagator $D(p) \sim p^{2(2\kappa_Z-1)}$ with infrared
exponents $\kappa_G = \kappa_Z \equiv \kappa \in [1/2,3/4]$. 
Note that $D(0) = 0$ for
$\kappa > 1/2$. The second solution, called {\em massive} or
{\em decoupling} solution, is characterized by a tree-level-like
ghost propagator at small momenta $G(p) \sim p^{-2}$ and by a
finite nonzero gluon propagator $D(0) > 0$ at zero momentum
(i.e., $\kappa_G = 0$, $\kappa_Z = 1/2$).

These two solutions have indeed been obtained by several groups
\cite{von Smekal:1997vx,Zwanziger:2001kw,Pawlowski:2003hq,
Aguilar:2004sw,Boucaud:2006if,Fischer:2006ub,Aguilar:2008xm}, using
different truncation schemes. Let us recall that the conformal solution
has also been obtained (in Landau gauge) in 2d and 3d Yang-Mills
theories \cite{Zwanziger:2001kw,Huber:2007kc} with the exponent
$\kappa$ approximately given by $\, 0.2(d-1)$, where $d=2,3$ and $4$ is
the space-time dimension. Possible explanations of the origin of these
two solutions have been discussed in \cite{Boucaud:2008ji,Boucaud:2008ky,
Fischer:2008uz}. For a description of confinement based on the conformal
solution see \cite{Alkofer:2000wg,Alkofer:2006fu}. On the
contrary, in \cite{Chaichian:2006bn} the authors relate massive
(respectively massless) gluons to color confinement (respectively
deconfinement). Color confinement in the presence of massive gluons
has also been related to the condensation of vortices \cite{Cornwall:1979hz}.
It is interesting to note that a criterion for quark confinement obtained
in Ref.\ \cite{Braun:2007bx} is satisfied both by the scaling and by the
massive solution.

The massive solution can also be obtained in 3d and 4d \cite{Dudal:2007cw,
Dudal:2008sp,Dudal:2008rm} by using the Gribov-Zwanziger approach. In this case
this solution appears when a suitable mass term is added to the action (while
preserving its renormalizability). In particular, the massive behavior is
related to the condensation of a mass-dimension $d-2$ operator. By setting to
zero the value of this condensate, one gets back the conformal solution. It is
interesting that the same approach cannot be extended to the 2d case
\cite{Dudal:2008xd}. Finally, the massive solution is also found by considering
a mapping (in the IR limit) of the Yang-Mills action onto the $\lambda \phi^4$ 
theory \cite{Frasca:2007uz}.


\subsection{Lattice studies}

Lattice simulations allow a true first-principles study of the
IR sector of QCD, with no uncontrolled approximations. However,
when studying the IR behavior of Green's functions in a given gauge,
care is needed in order to control possible finite-size, Gribov-copy
and discretization effects.

Recent numerical results for Landau gauge at very large
lattice volumes \cite{Cucchieri:2007md,Bogolubsky:2007ud,Sternbeck:2007ug}
indicate a finite gluon propagator $D(p)$ at zero momentum and a tree-level-like
IR ghost propagator $G(p)$ in 3d and in 4d. In particular, a flat ghost
dressing function --- or, equivalently, a null exponent $\kappa_G$ --- has been 
seen first in \cite{Oliveira:2007dy,Cucchieri:2007md,Bogolubsky:2007ud} and
more recently in \cite{Cucchieri:2008fc,Bornyakov:2008yx,Bogolubsky:2009dc}. 
A gluon propagator with an IR exponent
$\kappa_Z = 0.5$ has also been obtained using a tadpole-improved anisotropic
lattice action \cite{Gong:2008td}. On the contrary, in the 2d case, one
finds \cite{Maas:2007uv,Cucchieri:2007rg,Cucchieri:2008fc} $D(0)=0$ and
$G(p) \sim p^{-2(1+\kappa_G)}$ with $\kappa_G\approx \kappa_Z $ 
between 0.1 and 0.2. Note that this implies an
IR exponent $\kappa$ in relatively good agreement with the prediction
of the conformal solution \cite{Huber:2007kc}, i.e.\ $\kappa = 0.2$.
Thus, lattice simulations suggest a massive solution in 3d and in 4d,
while the 2d case seems consistent with the conformal solution. It is
important to note, however, that in the three cases one finds a clear
violation of reflection positivity \cite{Langfeld:2001cz,Cucchieri:2004mf,
Oliveira:2006zg,
Bowman:2007du} for $D(x)$ at $x \approx 1 fm$.
Also, lattice data confirm that, in the limit of large lattice volumes $V$,
the measure of the functional integration gets concentrated on the boundary
of the first Gribov region, in agreement with the Gribov-Zwanziger
confinement scenario \cite{Gribov:1977wm,Zwanziger:1993dh,Cucchieri:2008yp}.
Indeed, in this limit the smallest nonzero eigenvalue of the Faddeev-Popov
operator goes to zero \cite{Maas:2007uv,Cucchieri:1997ns,Sternbeck:2005vs,
Cucchieri:2006tf} in 2d, 3d and 4d Landau gauge.

The extrapolation of the gluon- and ghost-propagator data to the
infinite-volume limit has been recently improved by considering rigorous
upper and lower bounds \cite{Cucchieri:2007rg,Cucchieri:2008fc,
Cucchieri:2008mv} for $D(0)$ and $G(p_s)$, where $p_s$ is the smallest nonzero
momentum on the lattice. These bounds are valid at each lattice volume $V$ and
must be extrapolated to infinite volume, just as for the propagators.
However, the bounds are written in terms of quantities that are easier to
compute, better behaved or more intuitive than the propagators themselves.
This allows a more precise extrapolation and may provide a clearer
interpretation of the behavior of the propagators in terms of statistical 
averages. We note that similar bounds can also be written for $D(p)$ and $G(p)$
at general lattice momentum $p$, and for various gauge-fixing conditions.
Thus, the bounds can be used to check the necessary conditions for the 
IR-enhancement of $G(p)$ and for the IR-suppression of $D(0)$, clarifying 
when a Gribov-Zwanziger-like confinement scenario can be considered for a
given gauge \cite{Cucchieri:2006hi}.

Gribov-copy effects on gluon and ghost propagators have been extensively
studied on the lattice \cite{Cucchieri:1997dx,Silva:2004bv,
Silva:2007tt,Maas:2008ri,Bornyakov:2008yx}. Recent results
\cite{Bornyakov:2008yx}, using an extended
gauge-fixing procedure, suggest that the restriction of the configuration space 
to the so-called fundamental modular region $\Lambda$ \cite{Zwanziger:1993dh} 
produces a slightly more IR-suppressed gluon propagator. At the same time, all 
studies agree that this
restriction makes the ghost propagator less singular. This result has a very
simple explanation \cite{Cucchieri:1997dx} if one recalls that the fundamental modular
region $\Lambda$ is a subset of the first Gribov region $\Omega$ and that the
smallest nonzero eigenvalue $\lambda_{min}$ of the Faddeev-Popov matrix ${\cal M}$
is larger for configurations belonging to $\Lambda$ than for configurations in
$\Omega$ \cite{Cucchieri:1997ns}. Then, since $G(p) \sim {\cal M}^{-1}$ and $\lambda_{min}$
goes to zero in the infinite-volume limit, it is natural that $G(p)$ should be
smaller when a restriction to the fundamental modular region is implemented,
i.e.\ for any finite lattice volume we must have $G_{\Lambda}(p) < G_{\Omega}(p)$, where
the subscripts indicate the region considered for the functional integration.
Of course, after taking the infinite-volume limit one would still find
$G_{\Lambda}(p) \leq G_{\Omega}(p)$.

Finally, discretization effects are important for the breaking of
rotational symmetry as well as for the possible different discretizations
of the gluon field and of the gauge-fixing condition. In order to reduce
effects due to the breaking of rotational symmetry, three different
approaches have been considered: to cut out the momenta characterized by large
effects \cite{Leinweber:1998uu} (the so-called cylindrical cut),
to improve the lattice definition of the momenta \cite{Ma:1999kn} or to include
(hypercubic) corrections into the momentum-dependence of the Green's functions
\cite{deSoto:2007ht,Shintani:2008ga}. As for the discretization of the gluon field
and of the lattice Landau gauge condition, several different definitions can be
considered \cite{Langfeld:2001cz,DiRenzo:1997bb,Giusti:1998ur,Furui:1998cg,Bonnet:1999mj,
Cucchieri:1999dt,vonSmekal:2007ns}. These studies have usually found that different
discretization procedures lead to gluon propagators that differ only by a
multiplicative constant \cite{Giusti:1998ur,Cucchieri:1999dt,Bloch:2003sk}.
The situation may be different for the $\beta = 0$ case, discussed below.


\subsection{The case $\beta = 0$}

The case $\beta = 0$ corresponds to the (unphysical) strong-coupling limit
of lattice gauge theory. However, when studying the IR behavior of Green's
functions in a given gauge, it has some interesting advantages compared to
the usual simulations in the so-called scaling region. Indeed, at $\beta = 0$
the partition function determines that the gluon configuration should be completely
random, just as a spin model at infinite temperature. Thus, correlation functions
are probing only the effects due to the gauge-fixing condition. Moreover,
since at $\beta=0$ the lattice spacing is infinite (see Section \ref{sec:beta0}
below), any lattice volume considered is also infinite and we can hope for an
easier study of the infinite-volume limit of the theory. Finally, one should
also recall that the inequalities obtained in \cite{Zwanziger:1990by,
Dell'Antonio:1989jn} for the Landau-gauge gluon field and gluon propagator
are valid for any value of $\beta$, including $\beta = 0$.

Early numerical studies of the SU(2) gluon propagator at small $\beta$, 
including $\beta = 0$,
were presented in \cite{Cucchieri:1997dx,Cucchieri:1997fy}, showing the first
numerical evidence of a gluon propagator suppressed in the IR limit. In particular,
it was seen that, for $\beta=0$, the gluon propagator is decreasing (roughly
monotonically) as the momentum $p$ decreases. At small positive $\beta$,
the gluon propagator
also decreases for $p$ below a certain turn-over value $p_{to}$, with $p_{to}$
depending on the value of $\beta$. For this study the largest lattice volume 
considered was $V = 30^4$.
We note again that no such behavior is observed in the scaling region
on symmetric 4d lattices.


\subsection{Recent results at $\beta = 0$}
\label{sec:recent0}

A recent and extensive study at $\beta=0$, for the SU(2) case and
considering lattice volumes up to $56^4$, has been reported in
\cite{Sternbeck:2008wg, Sternbeck:2008mv,Sternbeck:2008na}.
In this case, the authors justify
the consideration of the case $\beta=0$ because it should correspond to
the formal limit of $\Lambda_{\rm QCD} \to \infty$. This, in turn, would allow
a study of the behavior of Green's functions for a range of momenta $\pi / L
\ll p \ll \Lambda_{\rm QCD}$, which is a necessary condition for the observation
of infrared behavior on the lattice.

Quoting the Conclusions of Ref.\ \cite{Sternbeck:2008mv},
the authors find that:
\begin{itemize}
\item the propagator's dressing functions show conformal
      scaling behavior for large lattice momenta, $a^2 q^2 \gg 1$;
\item finite-size effects are small;
\item the combined gluon and ghost data are consistent with
      an IR exponent $\kappa = 0.57(3)$;
\item both propagators show massive behavior at small momenta,
      i.e.\ $a^2 q^2 \ll 1$;
\item this massive behavior depends strongly on the lattice
      discretization used for the gluon field and for the
      gauge fixing;
\item while it is possible that this ambiguity disappears in
      the continuum limit, it is definitely present at commonly
      used values of the lattice couplings in SU(2).
\end{itemize}

We will comment on the analysis of the data in Section \ref{sec:plots}.
Here we would like to stress that we do not agree with the limit
$\Lambda_{\rm QCD} \to \infty$ as a motivation for this type of study.
First of all, the authors of \cite{Sternbeck:2008mv} do not quote
a value for the lattice spacing at $\beta = 0$ or for $\Lambda_{\rm QCD}$. 
Thus, it is
not clear how the comparison is really made. Of course, fixing the
lattice spacing in the strong-coupling regime is not a simple issue
(we will comment again on this issue in Section \ref{sec:beta0}). Indeed,
since we are far away from the continuum limit, the use of an experimental input
is not really justified. Nevertheless, if for example we fix $a$ from the physical
value of the string tension $\sigma$, then the strong-coupling
expansion [in the $SU(2)$ case] gives \cite{Creutz:1980zw} $a^2 \sigma = -
\log{(- \beta/4)}$ and in the limit $\beta \to 0$ we find $a \to \infty$.
In this case {\bf all masses}, evaluated on the lattice at $\beta = 0$, will have a
null physical value if expressed in physical units and a comparison to the
continuum physical value of $\Lambda_{\rm QCD}$ would be essentially meaningless,
since even the mass of a very heavy hadron would be infinitely small compared
to $\Lambda_{\rm QCD}$. One could also try to set a finite value for the lattice
spacing, as done for example in Ref.\ \cite{deForcrand:2009dh} using the
results presented in \cite{deForcrand:2006gu}. Then, depending on the mass
used as a physical input, one finds a lattice spacing ranging from $0.455 fm$
to $1.4 fm$. This corresponds to an inverse lattice spacing $1/a$ between 141
and $433 MeV$. Then, assuming for example a value for $\Lambda_{\rm QCD}$ of about
$225 MeV$ --- which corresponds to $\Lambda_{\rm \overline{MS}}$ --- we find that
$\Lambda_{\rm QCD}$ would have a value between 0.62 and 1.92 in lattice units 
at $\beta=0$. Thus, in this case, the condition $p \ll \Lambda_{\rm QCD}$ is not satisfied
by almost all the momenta considered in this work and in Refs.\ \cite{Sternbeck:2008mv,
Sternbeck:2008wg,Sternbeck:2008na}. Moreover, if one selects
as significant for the IR limit only data verifying this inequality, then
the data region with scaling behavior (described in the first item above)
should be discarded.

As for the ambiguity in the results related to different discretizations, it
seems to us that it is not really present if one considers all data already
available \footnote{Data have been extracted from the plots using
the {\em g3data} program.}. Indeed, in Fig.\ 11 of Ref.\ \cite{Sternbeck:2008mv},
the difference between the standard and the modified Landau gauge for the running coupling
constant when $\beta$ is 2.3 is about $0.84/0.73 \approx 1.15$ at the smallest
momentum (corresponding to $p^2 \approx 0.048 \, GeV^2$), for which the effect
seems to be larger. This is quite a large difference
but, as we will see below, it is probably mostly due to the ghost
sector. Indeed, the ghost dressing function enters quadratically into the definition
of the running coupling, i.e.\ this effect is artificially enlarged by
considering the running coupling instead of the propagators.
This difference decreases at larger momenta, being about 12.5\% at
$p^2 \approx 0.060 \, GeV^2$, 10\% at $p^2 \approx 0.222 \, GeV^2$ and 4.5\% at
$p^2 \approx 0.518 \, GeV^2$. (Here we used $a = 0.83814 \, GeV^{-1}$ in order to
convert lattice momenta to physical units.)
Now, if one looks at the data reported in Figure 1 of Ref.\ \cite{vonSmekal:2007ns}
for the case $\beta = 2.5$, there are ``hardly any differences'' between
the two different discretizations, as stressed by
the authors themselves. To be more precise, the difference in the gluon field
is clearly within error bars, while for the ghost dressing function the ratio
between the two results seems close to $2.06/2.02 \approx 1.025$, again considering
the smallest nonzero momentum (corresponding to $p^2 \approx 0.206 \, GeV^2$).
This implies a difference of about 5\% for the running coupling (to be compared
to a 10\% difference at a similar value of $p^2$ for $\beta = 2.3$). For the
next point, corresponding to $p^2 \approx 0.413 \, GeV^2$, this difference is
about 2.4\%. Thus, the discretization effects observed in Ref.\
\cite{Sternbeck:2008mv} seem to disappear in the continuum limit, being already reduced
by a factor 2 when going from $\beta = 2.3$ to $\beta = 2.5$. Of course, a volume
$V = 32^4$ at $\beta = 2.5$ is rather small. On the other hand, data for the gluon and
the ghost propagators at $p \approx 0.4-0.6 \, GeV$, usually do not
suffer from large finite-size effects \cite{Bloch:2003sk}. One should also
recall that large discretization effects in the ghost sector when using the
so-called modified Landau gauge can be easily explained: indeed, in this case
the discretized Faddeev-Popov matrix has an extra term, which is quadratic in
the gluon field \cite{vonSmekal:2007ns,Sternbeck:2008mv}. This term, of course,
is not present in the continuum expression for the Landau Faddeev-Popov term and
is also not present in the standard lattice Landau gauge. Thus, it is very 
plausible that the large
discretization effects observed in \cite{Sternbeck:2008mv} be due
to this peculiar characteristic of the modified Landau gauge. This could also explain why
in this case discretization effects cannot be accounted for by a global multiplicative
factor as in \cite{Giusti:1998ur,Cucchieri:1999dt,Bloch:2003sk}. 
By considering all this, we do not find the discretization effects studied 
in \cite{Sternbeck:2008mv} to be of significant relevance in the analysis 
of the IR behavior
of Green's functions in Landau gauge. In any case, we are now studying these
discretization effects in more detail \cite{preparation}.

Finally, we would like to stress that while simulations at null $\beta$ may be
interesting for a qualitative analysis \cite{Smit}, they should clearly 
not be taken too seriously at the quantitative level. Thus, a value of
$\kappa$ close to the preferred value of the so-called scaling solution ---
see the third item above --- should at most be considered as a peculiar
coincidence, not as a physically relevant result.


\section{Numerical simulations and results}
\label{sec:num}

Here we present data for the gluon and the ghost propagators in Landau gauge
at $\beta = 0$ both in 3d and in 4d. The 3d data, with lattice volumes up to
$100^3$, have been presented in \cite{IRQCDRio}. The analysis presented here
and the 4d data, at $V = 64^4$, are new. We consider the $SU(2)$ case.
Let us recall that recent numerical studies 
\cite{Cucchieri:2007zm,Sternbeck:2007ug,Cucchieri:2007ji,Maas:2007af,Cucchieri:2007bq}
have verified that IR Landau-gauge gluon and ghost propagators are rather similar
for the SU(2) and the SU(3) gauge groups, as expected from DSE studies. 
Thus, the analysis presented here is likely valid for the SU(3) case as well.

In the 3d case we considered seven lattice volumes, namely $V = 10^3,
20^3, 30^3, 40^3, 60^3, 80^3, 100^3$, with (respectively) 1000, 900,
773, 700, 1240, 344 and 364 configurations. In the 4d case we produced
data only for the lattice volume $V = 64^4$, with a total of 567
configurations. The gauge fixing has been
done using the stochastic-overrelaxation method \cite{Cucchieri:1995pn}.
For the ghost propagator, due to technical reasons related to the
computers used for the simulations, we employ the point-source method
described in \cite{Cucchieri:2006tf,Boucaud:2005gg}.
Let us note that the use of the point-source method usually
increases the statistical noise in the evaluation of the ghost
propagator as compared to the plane-wave source \cite{Cucchieri:2006tf}.
However, in the 4d case with more than 550 configurations we have that
more than 70\% of the data have a relative
error smaller than 3\% and 94.5\% of the data have a relative error smaller
than 5\%. So, the introduced fluctuations are clearly compensated by our
increased statistics.

For our study of the gluon propagator in the 3d case we considered data for
three different types of momenta, i.e.\ momenta with components $(0,0,q)$,
$(0,q,q)$ and $(q,q,q)$, plus possible permutations of the Lorentz index.
This implies that data corresponding to the momenta $(0,0,q)$ and $(0,q,q)$
have a statistics three times as large as the data corresponding to the momenta
$(q,q,q)$. In the ghost case we evaluate the propagator only for the momenta
$(0,0,q)$ and $(0,q,q)$ (plus permutations of the Lorentz index). Similarly,
in 4d we consider momenta with components $(0,0,0,q)$, $(0,0,q,q)$, $(0,q,q,q)$
and $(q,q,q,q)$. For the ghost (respectively gluon) propagator we fully (respectively
partially) applied permutation of the Lorentz index.


\subsection{Lattice spacing at $\beta = 0$}
\label{sec:beta0}

As shown above, it is not simple to fix the lattice spacing in the strong-coupling
regime. Indeed, if one uses an experimental input the result varies from a finite
large value to an infinite value. This uncertainty is of course a manifestation
of the possible discretization effects encountered at $\beta=0$.

Nevertheless, we believe that for the study of correlation
functions in Landau gauge at $\beta=0$ one should consider the lattice spacing
as infinite. This result does not use an experimental input but it is based
on the properties of the gauge-fixing algorithms used for fixing the random
configurations to Landau gauge. Since all the ``dynamics'' in the present
study is produced by the gauge-fixing procedure, it seems to us more reliable to use
the gauge-fixing process as an input for fixing the lattice spacing.
To this end we notice that if $a = \infty$ then
all lattice volumes correspond to infinite lattice size (in physical units)
and all (non-zero) correlation lengths are also infinite (again in physical
units). This means that at $\beta=0$ we can work at {\em constant physics} --- i.e.\
fixed correlation length in physical units --- by simply changing the lattice volume
\cite{Cucchieri:2003fb} without tuning the value of $\beta$, as one has to do when
using finite nonzero values of $\beta$. (The same is true at $\beta = \infty$,
where all lattice sides and correlation lengths are zero in physical units.) Indeed,
the evaluation of critical slowing-down of standard Landau gauge-fixing algorithms,
which requires working at constant physics \cite{Cucchieri:1995pn}, can be done also
at $\beta = 0$ \cite{Cucchieri:1996jm}, yielding the expected critical exponents.

Of course, since we assume $a = \infty$, in this work all data will be presented
in lattice units.


\begin{figure}[t]
\vspace{-43mm}
\includegraphics[scale=0.48]{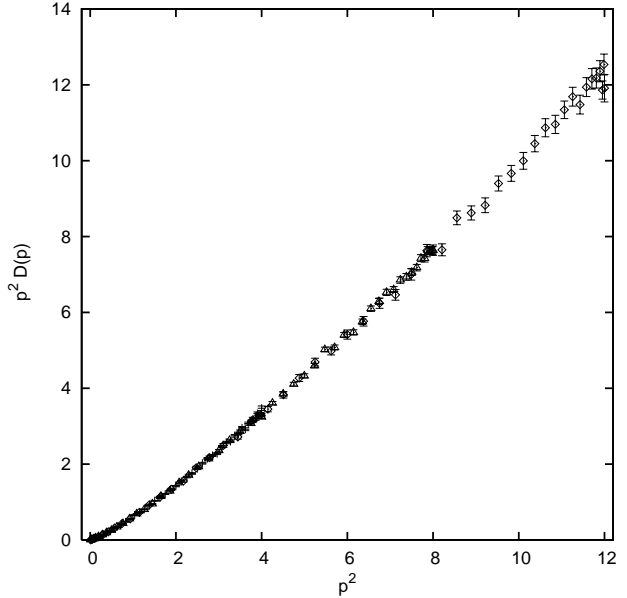}
\caption{\label{fig:Z-100}
  The gluon dressing function $p^2 D(p)$ as a function of the (unimproved)
  lattice momenta $p^2$ for the lattice volume $V = 100^3$. Symbols $+$,
  $\triangle$ and $\Diamond$ represent data corresponding (respectively) to momenta
  $(0,0,q)$, $(0,q,q)$ and $(q,q,q)$. Recall that for these types of momenta the
  largest value of $p^2$ is respectively equal to 4, 8 and 12.
  }
\end{figure}

\subsection{Breaking of rotational invariance}
\label{sect:breaking}

In the continuum, the gluon propagator $D(p)$ and the ghost propagator $G(p)$
are just functions of $p^2 = \sum_{\mu} p_{\mu}^2$. On the lattice, the
momentum components $p_{\mu}$ are usually given by $p_{\mu} = 2 \sin(\pi n_{\mu}/
N_{\mu})$, where $n_{\mu}$ is an integer in the interval $1-N_{\mu}/2, \ldots,
N_{\mu}/2$ and $N_{\mu}$ (here supposed even) represents the number of sites in
the $\mu$ direction. However, as explained in the Introduction, due to the breaking
of rotational symmetry, one expects the gluon and ghost propagators on the
lattice to depend also on hypercubic corrections \cite{Ma:1999kn,deSoto:2007ht,
Shintani:2008ga}, with the leading term proportional to $p^{[4]} =
\sum_{\mu} p_{\mu}^4$. In this Section we try to quantify these corrections and
see if they can introduce systematic effects on the analysis of the data. 
Note that, to this end, we do not need to consider the $\beta$-dependence
of these corrections,
as described in Section \ref{sec:recent0} above for the discretization errors of the
gluon field, but only estimate the corrections at the considered value of $\beta$.

In order to get rid of these corrections one should
extrapolate the numerical data to the limit $p^{[4]} \to 0$. This can be
done when data are available for momenta characterized by the same value
of $p^2$ but with different values of $p^{[4]}$, i.e.\ they belong to the same
$O(4)$ orbit but to different $H(4)$ orbits (see \cite{deSoto:2007ht} for
description and notation). With the choice of momenta considered in the 3d 
case here (see the third paragraph of Section \ref{sec:num} above) the 
extrapolation $p^{[4]}
\to 0$ cannot be easily done, since we essentially have only one $H(4)$ orbit
for each $O(4)$ orbit. The situation does not improve if one uses a different
discretization for the lattice momenta \cite{deSoto:2007ht}, e.g.\ $p_{\mu}
= 2 \pi n_{\mu}/ N_{\mu}$.

As said in the Introduction, rotational symmetry can also be (partially) restored
by considering \cite{Ma:1999kn} an improved lattice momentum with magnitude
\begin{equation}
p^2_i = p^2 \left[ 1 + s \, p^{[4]} / p^2 \right] \; ,
\label{eq:pimpr}
\end{equation}
where $s$ is a numerical coefficient. (When considering physical units,
$s$ has mass dimension $-2$, i.e.\ it is proportional to $a^2$.)
Note that, with the momenta considered here in
the 3d case and for a fixed momentum $p^2 \neq 0$, the correction $p^{[4]} / p^2$ is
maximal, and equal to $p^2$, if the momentum $p$ has only one component different
from zero. On the contrary, this correction is minimal, and equal to $p^2 / d$
(where $d$ is the space-time dimension), if all the components of $p$ are equal.
Also note that the cylindrical cut \cite{Leinweber:1998uu} selects data for which
this correction is relatively small, i.e.\ data close to the diagonal direction
$(q,q,q)$.

In order to estimate the value of the coefficient $s$ in Eq.\ (\ref{eq:pimpr}), we
consider 3d gluon data corresponding to a certain type of momenta, e.g.\ $(0,q,q)$. Then,
for a given value of $s$, we find a spline describing these data as a function of
the improved lattice momentum $p^2_i$. Finally, we use this spline as a fitting function
for the data corresponding to the other two types of momenta --- $(0,0,q)$ and $(q,q,q)$
in this example --- again considered as a function of the improved lattice momentum $p^2_i$,
with the same value of $s$. Let us recall that a spline is usually quite unstable
outside the set of data that it describes. Thus, this fit has been done only for data
corresponding to a value of $p^2_i$ that is inside the range of momenta used for the
evaluation of the spline. Also, the zero-momentum datum has not been used for this analysis.
Clearly, the chosen value of $s$, for each lattice volume, is the one that minimizes the
$\chi^2 / dof$ value for the fit. We find that the coefficient $s$ obtained in this
way depends on the type of momenta used to evaluate the spline. This is mainly related to
the fact that the ranges described by the three different types of momenta are quite
different. However, in most cases we found a value for $s$ of the order of $\approx 0.01-0.02$
and in all cases we found values smaller than the perturbative value $s = 1/12 \approx 0.0833$
\cite{Ma:1999kn}, which is usually a good guess for data in the scaling region (see for
example the ghost propagator in maximally Abelian gauge \cite{Mendes:2008ux}).
That the effects of breaking rotational invariance are small is also confirmed (see Fig.\
\ref{fig:Z-100}) by the plot of the gluon dressing function. Note that these effects are
usually more visible when considering $p^2 D(p)$ instead of $D(p)$ \cite{Becirevic:1999uc}.
The situation is similar for the ghost propagator in the 3d case and for the gluon and ghost
propagators in the 4d case.

One should note that having small rotational-symmetry-breaking effects at $\beta = 0$
is perhaps a surprising result, since one might expect strong discretization errors
given that $a$ is infinite. This is probably due to the fact that at $\beta = 0$ all
data are effectively in the deep IR region and in this limit violation of rotational
symmetry is usually quite small.


\subsection{Plots and fits}
\label{sec:plots}

In order to test for behavior according to the conformal or to the decoupling 
solutions --- and in analogy with
Refs.\ \cite{Sternbeck:2008wg,Sternbeck:2008mv,Sternbeck:2008na} --- we tried a fit
of the numerical data using for the gluon propagator the fitting function
$b+c (p^2)^{2 \kappa_Z+1-d/2}$, where $d$ is the number of space-time dimensions, 
and using for the ghost propagator the fitting function $c (p^2)^{-\kappa_G-1}$.
For the gluon propagator we considered the cases $b = 0$ and $b \neq 0$,
corresponding to Eqs.\ (18a) and (18b) of Ref.\ \cite{Sternbeck:2008mv}. The
function used for the ghost corresponds to Eq.\ (19a) of the same reference.
However, note that we usually fit the ghost propagator and not the ghost dressing
function as in Ref.\ \cite{Sternbeck:2008wg} (we will comment again on this point
in Section \ref{sec:massive} below). Also recall
that the scaling solution \cite{Huber:2007kc} gives $D(p) \sim (p^2)^{2 \kappa+1-d/2}$
and $G(p) \sim (p^2)^{-\kappa-1}$, with $\kappa \approx 0.3976$ in 3d and $\kappa \approx
0.5953$ in 4d. The fits have also been done separately for the different types of
momenta used in our simulations. This allows again an estimate of systematic effects
due to the breaking of rotational invariance.


\subsubsection{3d gluon propagator} 
\label{sec:3dgluon}

\begin{table}
\vskip -2mm
\caption{Parameter fit for the gluon propagator $D(p)$ as a function of
         the (unimproved) lattice momentum $p^2$ using the fitting function
         $b+c (p^2)^{2 \kappa_Z-0.5}$ and data points in the range $p^2 \in
         [0,1.5]$. We do a separate fit for each of the three types of momenta
         $(0,0,q)$, $(0,q,q)$ and $(q,q,q)$, here indicated (respectively) as
         1,2 and 3. For each fit we also report the number of degrees of
         freedom ($dof$) of the fit and the value of $\chi^2/dof$. Notice that we considered
         only the lattice volumes $V \geq 40^3$ in order to have enough data points for
         the fit.
\label{tab:gluon-low}}
\vskip 2mm
\begin{tabular}{ccccccc}
\hline
 $V$     & momenta &   $b$    &   $c$    &   $\kappa_Z$  & $dof$ & $\chi^2 / dof$ \\
 \hline
 $40^3$  &    1    & 0.319(8) & 0.299(9) & 0.51(2)  & 6 & 2.38 \\
 $40^3$  &    2    & 0.28(3)  & 0.34(3)  & 0.47(4)  & 2 & 3.64 \\
 $40^3$  &    3    & 0.33(5)  & 0.30(5)  & 0.5(1)   & 1 & 3.22 \\
 \hline
 $60^3$  &    1    & 0.308(6) & 0.307(6) & 0.50(1)  & 10 & 2.40 \\
 $60^3$  &    2    & 0.29(1)  & 0.33(1)  & 0.48(2)  & 5 & 1.90 \\
 $60^3$  &    3    & 0.23(6)  & 0.37(6)  & 0.42(5)  & 3 & 3.13 \\
 \hline
 $80^3$  &    1    & 0.314(6) & 0.307(6) & 0.52(1)  & 14 & 1.50 \\
 $80^3$  &    2    & 0.29(1)  & 0.32(1)  & 0.48(2)  & 8 & 1.37 \\
 $80^3$  &    3    & 0.309(9) & 0.296(9) & 0.51(2)  & 6 & 0.24 \\
 \hline
 $100^3$ &    1    & 0.305(5) & 0.307(6) & 0.50(1)  & 18 & 1.73 \\
 $100^3$ &    2    & 0.28(1)  & 0.33(1)  & 0.46(1)  & 11 & 1.61 \\
 $100^3$ &    3    & 0.30(1)  & 0.32(1)  & 0.49(2)  & 8 & 0.97 \\
 \hline
\end{tabular}
\vskip 1mm
\end{table}

\begin{table}
\vskip -2mm
\caption{Parameter fit for the gluon propagator $D(p)$ as a function of
         the (unimproved) lattice momentum $p^2$ using the fitting function
         $c (p^2)^{2 \kappa_Z-0.5}$ and data points in the range $p^2 \geq 1.5$.
         We do a separate fit for each of the three types of momenta
         $(0,0,q)$, $(0,q,q)$ and $(q,q,q)$, here indicated (respectively) as
         1,2 and 3. For each fit we also report the number of degrees of
         freedom ($dof$) of the fit and the value of $\chi^2/dof$.
\label{tab:gluon-high}}
\vskip 2mm
\begin{tabular}{ccccccc}
\hline
 $V$     & momenta &   $c$    &   $\kappa_Z$    & $dof$ & $\chi^2 / dof$ \\
 \hline
 $10^3$  &    1    & 0.65(2)  & 0.34(2)  &  1  &  1.43 \\
 $10^3$  &    2    & 0.640(3) & 0.349(2) &  2  &  0.11 \\
 $10^3$  &    3    & 0.65(4)  & 0.34(1)  &  2  &  2.77 \\
 \hline
 $20^3$  &    1    & 0.628(9) & 0.357(6) &  4  &  0.85 \\
 $20^3$  &    2    & 0.623(6) & 0.355(3) &  6  &  1.25 \\
 $20^3$  &    3    & 0.64(2)  & 0.344(7) &  6  &  2.11 \\
 \hline
 $30^3$  &    1    & 0.6291)  & 0.356(8) &   7  &  2.12 \\
 $30^3$  &    2    & 0.631(7) & 0.351(3) &   9  &  1.46 \\
 $30^3$  &    3    & 0.625(9) & 0.351(4) &  10  &  0.91 \\
 \hline
 $40^3$  &    1    & 0.637(7) & 0.348(5) & 10 & 1.15 \\
 $40^3$  &    2    & 0.620(4) & 0.356(2) & 13 & 0.61 \\
 $40^3$  &    3    & 0.64(1)  & 0.347(4) & 14 & 1.58 \\
 \hline
 $60^3$  &    1    & 0.625(5) & 0.355(4) & 16 & 1.23 \\
 $60^3$  &    2    & 0.620(3) & 0.355(2) & 20 & 0.70 \\
 $60^3$  &    3    & 0.618(7) & 0.355(3) & 22 & 1.65 \\
 \hline
 $80^3$  &    1    & 0.626(7) & 0.354(5) & 22 & 1.41 \\
 $80^3$  &    2    & 0.627(5) & 0.352(2) & 27 & 0.77 \\
 $80^3$  &    3    & 0.624(9) & 0.352(4) & 29 & 1.17 \\
 \hline
 $100^3$ &    1    & 0.616(5) & 0.362(4) & 28 & 1.06 \\
 $100^3$ &    2    & 0.625(4) & 0.354(2) & 34 & 0.87 \\
 $100^3$ &    3    & 0.629(6) & 0.349(2) & 37 & 0.62 \\
 \hline
\end{tabular}
\vskip 1mm
\end{table}

\begin{table}
\vskip -2mm
\caption{Parameter fit for the gluon propagator $D(p)$ as a function of
         the (unimproved) lattice momentum $p^2$ using the fitting function
         $b+c (p^2)^{2 \kappa_Z-0.5}$ and all data points. We do a separate fit for each
         of the three types of momenta $(0,0,q)$, $(0,q,q)$ and $(q,q,q)$,
         here indicated (respectively) as 1,2 and 3. For each fit we also
         report the number of degrees of freedom ($dof$) of the fit and the
         value of $\chi^2/dof$.
\label{tab:gluon-all}}
\vskip 2mm
\begin{tabular}{ccccccc}
\hline
 $V$     & momenta &  $b$  &   $c$     &    $\kappa_Z$    & $dof$ & $\chi^2 / dof$ \\
 \hline
 $10^3$  &  1  &  0.49(1)  &  0.16(1)  &  0.55(2)  &  3  &  2.09 \\
 $10^3$  &  2  &  -0.2(2)  &  0.9(2)   &  0.33(1)  &  2  &  0.17 \\
 $10^3$  &  3  &  -1(7)    &  2(7)     &  0.3(1)   &  2  &  2.74 \\
 \hline
 $20^3$  &  1  &  0.37(1)  &  0.24(2)  &  0.51(2)  &  8  &  7.84 \\
 $20^3$  &  2  &  0.18(5)  &  0.44(6)  &  0.39(2)  &  7  &  2.26 \\
 $20^3$  &  3  &  -0.1(2)  &  0.7(2)   &  0.34(3)  &  7  &  2.64 \\
 \hline
 $30^3$  &  1  &  0.33(1)  &  0.28(2)  &  0.48(2)  & 13  &  8.22 \\
 $30^3$  &  2  &  0.12(4)  &  0.50(4)  &  0.38(1)  & 12  &  2.42 \\
 $30^3$  &  3  &  0.09(6)  &  0.52(7)  &  0.37(1)  & 12  &  1.47 \\
 \hline
 $40^3$  &  1  &  0.30(1)  &  0.31(1)  &  0.46(1)  & 18  &  5.93 \\
 $40^3$  &  2  &  0.17(2)  &  0.44(2)  &  0.392(8) & 17  &  2.24 \\
 $40^3$  &  3  &  0.08(7)  &  0.54(7)  &  0.36(1)  & 17  &  2.75 \\
 \hline
 $60^3$  &  1  &  0.289(8) &  0.32(1)  &  0.457(7) & 28  &  5.42 \\
 $60^3$  &  2  &  0.21(1)  &  0.40(2)  &  0.405(6) & 27  &  3.48 \\
 $60^3$  &  3  &  0.17(3)  &  0.45(3)  &  0.387(8) & 27  &  2.35 \\
 \hline
 $80^3$  &  1  &  0.289(9) &  0.32(1)  &  0.455(8) & 38  &  3.59 \\
 $80^3$  &  2  &  0.22(1)  &  0.39(2)  &  0.406(6) & 37  &  2.39 \\
 $80^3$  &  3  &  0.20(3)  &  0.41(3)  &  0.398(9) &  37  &  1.79 \\
 \hline
 $100^3$ &  1  &  0.288(6) &  0.315(7) &  0.460(6) &  48  &  2.45 \\
 $100^3$ &  2  &  0.24(1)  &  0.37(1)  &  0.413(5) &  47  &  2.35 \\
 $100^3$ &  3  &  0.20(2)  &  0.42(2)  &  0.393(6) &  47  &  1.40 \\
 \hline
\end{tabular}
\vskip 1mm
\end{table}

\begin{figure}
\vspace{-43mm}
\includegraphics[scale=0.48]{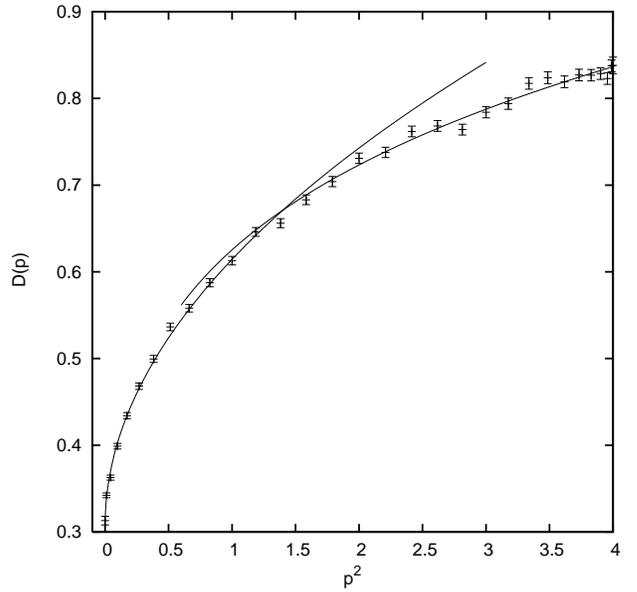}
\caption{\label{fig:3d-60}
  The gluon propagator $D(p)$ as a function of the (unimproved)
  lattice momenta $p^2$ for the lattice volume $V = 60^3$ for momenta of
  the type $(0,0,q)$. We also present the fits obtained at small momenta $p^2
  \leq 1.5$ and at large momenta $p^2 > 1.5$ (see Tables \ref{tab:gluon-low}
and \ref{tab:gluon-high}).
  }
\end{figure}

\begin{figure}
\vspace{-42mm}
\includegraphics[scale=0.48]{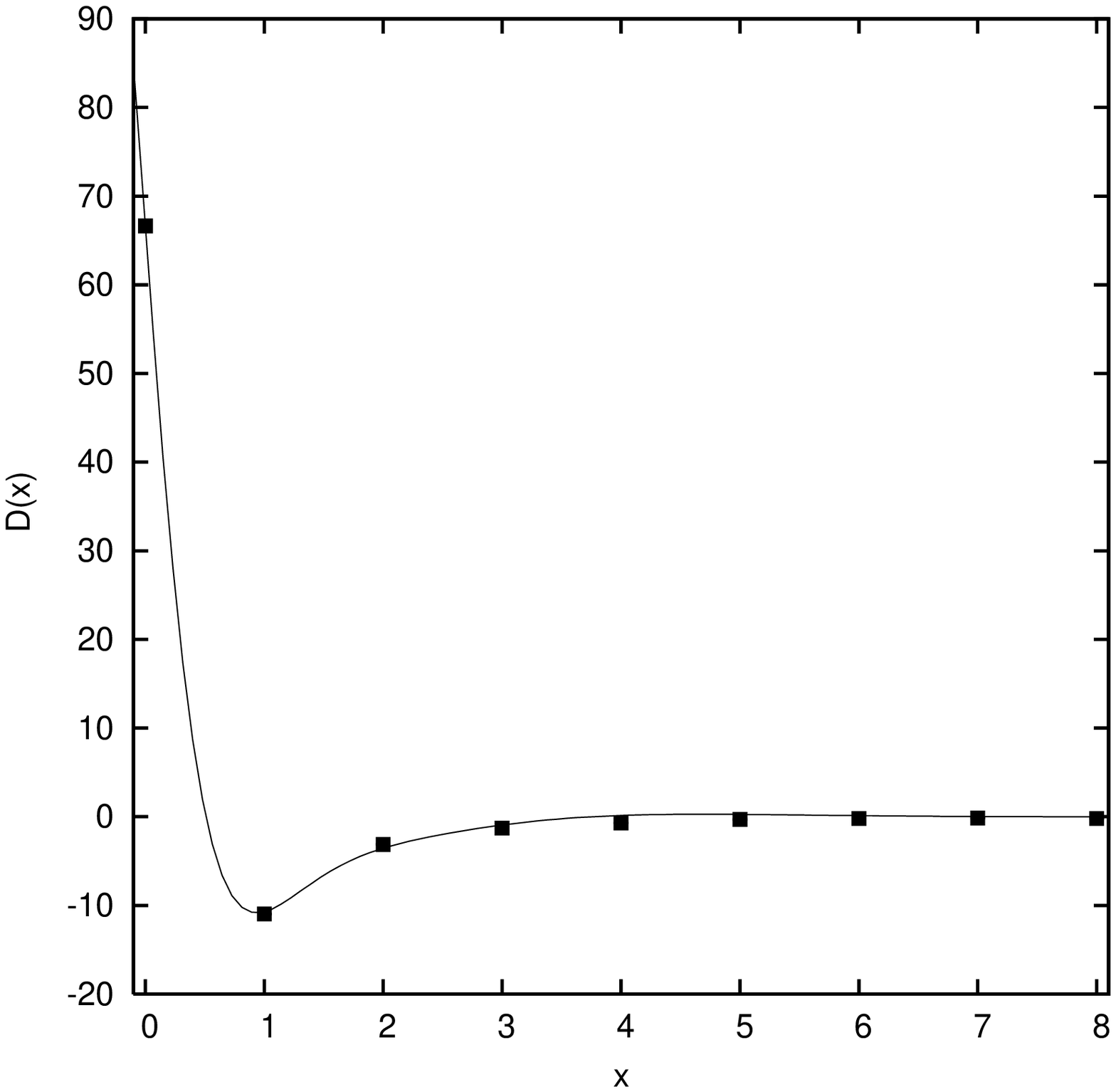}
\caption{\label{fig:3d-t-100}
  The gluon propagator $D(x)$ as a function of the space-time separation
  $x$ for the lattice volume $V = 100^3$, obtained by Fourier transforming
  data corresponding to momenta of the type $(0,0,q)$. For clarity, only the interval
  $x \in [0,8]$ is represented here. We also present the fit obtained
  using the fitting function $f(x)$ described in the text.
  }
\end{figure}

In the gluon case we find that data at small momenta are best fitted with $b\neq0$.
Then, the exponent $\kappa_Z$ is very close to 0.5 in the 3d case (see Table
\ref{tab:gluon-low}) if one uses for the fit the range $p^2 \in [0, p^2_{m}]$
with $p^2_{m} = 1.5$. A slightly larger or smaller value for $\kappa_Z$ is obtained
if one uses respectively $p^2_{m} = 1$ or $p^2_{m} = 2$. For momenta larger than
$p^2_{m}$ the best fit is obtained considering $b = 0$. In this case (see
Table \ref{tab:gluon-high}) the exponent $\kappa_Z$ is very close to 0.35 and this
value is essentially independent of the value of $p^2_{m}$. In both cases the
effects due to the breaking of rotational invariance are relatively small,
as expected (see discussion in Section \ref{sect:breaking} above). 
Finally, we tried a fit of all the data with $b\neq0$ (see Table
\ref{tab:gluon-all}). Clearly, in this case one finds for $\kappa_Z$ a kind of
average between the value 0.5 found at small momenta and the value 0.35 obtained
at large momenta. It is interesting to notice that a value very close to the 
conformal solution $\kappa \approx 0.3976$ is obtained for the largest lattice
volumes and for momenta of the type $(0,q,q)$ and $(q,q,q)$. As an example
we report in Fig.\ \ref{fig:3d-60} the fits obtained at small and at large
momenta for the lattice volume $V = 60^3$ and for data corresponding to the
momenta $(0,0,q)$.

One can also check that the gluon propagator $D(x)$ violates reflection positivity.
Actually, for all cases we find that $D(x=1)$ is already negative. [Of course, $D(x=0)$
is always positive.] In Fig.\ \ref{fig:3d-t-100} we show $D(x)$ as a function
of the space-time separation $x$ for the lattice volume $V = 100^3$ when considering
momenta of the type $(0,0,q)$. Following Ref.\ \cite{Cucchieri:2004mf}, we
fitted the data using a sum of two Stingl-like \cite{Stingl:1985hx} propagators
$f(x) = f_1(x) + f_2(x)$ with $f_i(x) = c_i \, \cos{(b_i + \lambda_i x)}
\, e^{-\lambda_i x}$.
We found a good description of the data (see again Fig.\ \ref{fig:3d-t-100}) by
fixing $c_1 = D(x=0) = 66.6393, b_1 = 0, b_2 = \pi/2, \lambda_2 = \lambda_1 / 3$
and with $c_2 = 20(2)$ and $\lambda_1 = 2.5(2)$. Also, by comparing these data
to results obtained in the scaling region \cite{Cucchieri:2004mf}, one can use the
value of $x$ where $D(x)$ starts to oscillate around zero as an input for
fixing the lattice spacing. From Fig.\ \ref{fig:3d-t-100} this happens at $x \approx
3$, giving $a \approx 1 fm$ (see Figure 5 in Ref.\ \cite{Cucchieri:2004mf}). This
is in quantitative agreement with the values obtained in \cite{deForcrand:2009dh}
(see Section \ref{sec:recent0} above).

\begin{figure}
\vspace{-44mm}
\includegraphics[scale=0.48]{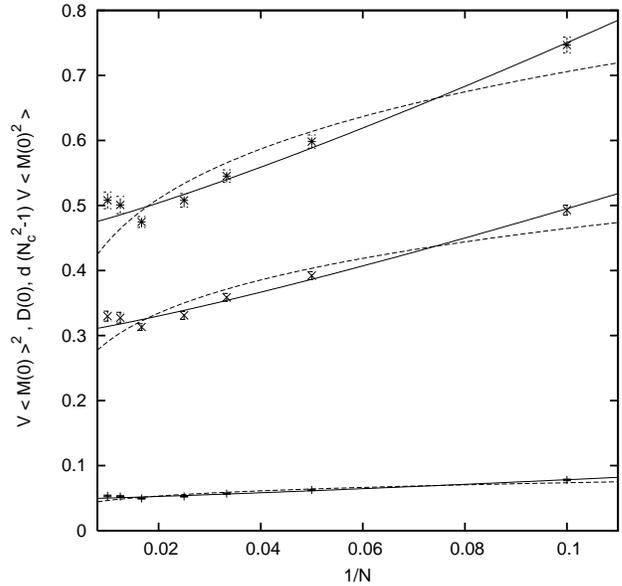}
\caption{\label{fig:bounds-D-3d-zero}
  The gluon propagator at zero momentum $D(p=0)$, as well as the upper and the
  lower bounds introduced in \cite{Cucchieri:2007rg}, as a function of the
  inverse lattice size $1/N$. We also present a fit to the data using the
  function $a+b/ N^c$ (solid line) and the function $e/ N^d$ (dashed line)
  (see Table \ref{tab:bounds}).
  }
\end{figure}

We also consider the volume dependence of $D(p=0)$ and the bounds introduced in
\cite{Cucchieri:2007rg}. As one can see from Fig.\ \ref{fig:bounds-D-3d-zero},
these bounds are well satisfied. Compared
to the results obtained at finite (nonzero) $\beta$ (see Fig.\ 1 in
\cite{Cucchieri:2007rg}), we find that the values of $D(0)$ are closer to the
upper bound than to the lower one. We have also checked that we can
extrapolate the data for $D(0)$ and for the upper and the lower bounds
to a finite nonzero value as well as to zero (see Table \ref{tab:bounds}).
However, considering the value of $\chi^2 / dof$, our data seem to prefer
an extrapolation to $D(0) \neq 0$ (see also Fig.\ \ref{fig:bounds-D-3d-zero}).

\begin{figure}
\vspace{-43mm}
\includegraphics[scale=0.48]{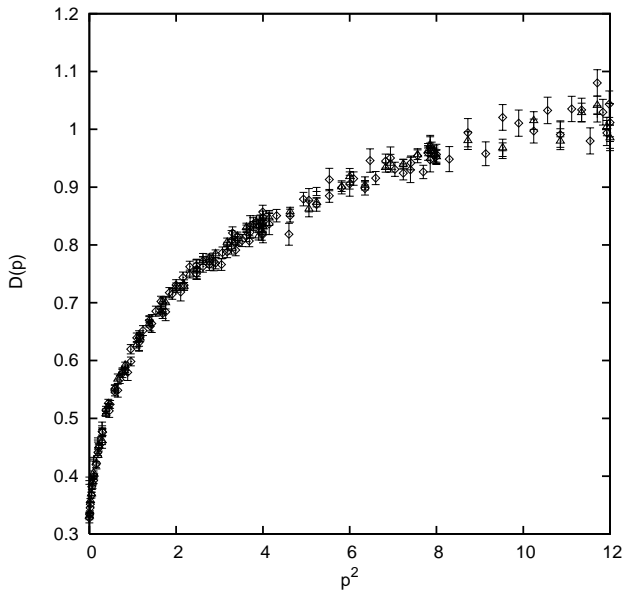}
\caption{\label{fig:3d-L}
  The gluon propagator $D(p)$ as a function of the (unimproved)
  lattice momenta $p^2$ for the lattice volumes $V = 20^3$ (symbol $+$),
  $40^3$ (symbol $\triangle$) and $80^3$ (symbol $\Diamond$).
  Here all types of momenta are represented for the three lattice volumes
  considered.
  }
\end{figure}

Let us note that the above result is in agreement with what observed in
Ref.\ \cite{Cucchieri:2003di}, i.e.\ in the 3d case volumes up to $100^3$
are not large enough in order to have a complete control over the
extrapolation to $V = \infty$ at $p=0$. This suggests that --- for the zero-momentum
propagator $D(0)$ --- simulations for $\beta = 0$ and for $\beta$ values in
the scaling region are essentially equivalent. In other words, the value of $D(0)$ 
seems to be related only to the thermodynamic limit ($V \to \infty$) and is not 
affected by the value of the lattice spacing. 
This peculiar behavior at $p=0$ may be related to the fact that the
gauge-fixing condition $p \cdot A(p) = 0$ does not play a role in this case. Also,
one might argue that $D(0)$ is simply a measure of the zero modes of the gluon field,
whose value is of course strongly affected by the lattice size.
The situation is different for $p \neq 0$.
Indeed, while at $\beta = 0$ the propagator is decreasing with $p$ for all lattice 
volumes and momenta, for values of $\beta$ in the scaling region
one sees a decreasing propagator only for large enough lattice volume and
for momenta below a certain value $p_{to}$.
Also, at $\beta = 0$ finite-size effects for $p > 0$ are very small
or null (see Fig.\ \ref{fig:3d-L} below and compare to Fig.\ 1 in 
Ref.\ \cite{Sternbeck:2008mv}), while for $\beta > 0$ one sees
large finite-size effects for $p \leq p_{to}$.

\begin{table}
\vskip -2mm
\caption{Parameter fit for the gluon propagator at zero momentum $D(p=0)$, as
         well as for the upper and the lower bounds, as a function of the inverse
         lattice size $1/N$ using the fitting functions $a+b / N^c$ and $e / N^d$.
         For each fit we also report the value of $\chi^2/dof$.
\label{tab:bounds}}
\vskip 2mm
\begin{tabular}{ccccc}
\hline
 fit   &   $a$   &   $b$    &   $c$    & $\chi^2 / dof$ \\
 \hline
 lower bound  & 0.048(3) & 0.5(4) & 1.2(4) & 4.22 \\
 $D(0)$       & 0.30(2)  & 3(2)   & 1.2(3) & 3.42 \\
 upper bound  & 0.46(3)  & 4(3)   & 1.2(3) & 3.62 \\
 \hline
 fit   &         &   $e$    &   $d$    & $\chi^2 / dof$ \\
 \hline
 lower bound  &          & 0.12(2) & 0.20(4) & 13.0 \\
 $D(0)$       &          & 0.74(9) & 0.20(4) & 13.5 \\
 upper bound  &          & 1.1(1)  & 0.20(4) & 12.3 \\
 \hline
\end{tabular}
\vskip 1mm
\end{table}

\subsubsection{4d gluon propagator} 

\begin{figure}
\vspace{-43mm}
\includegraphics[scale=0.48]{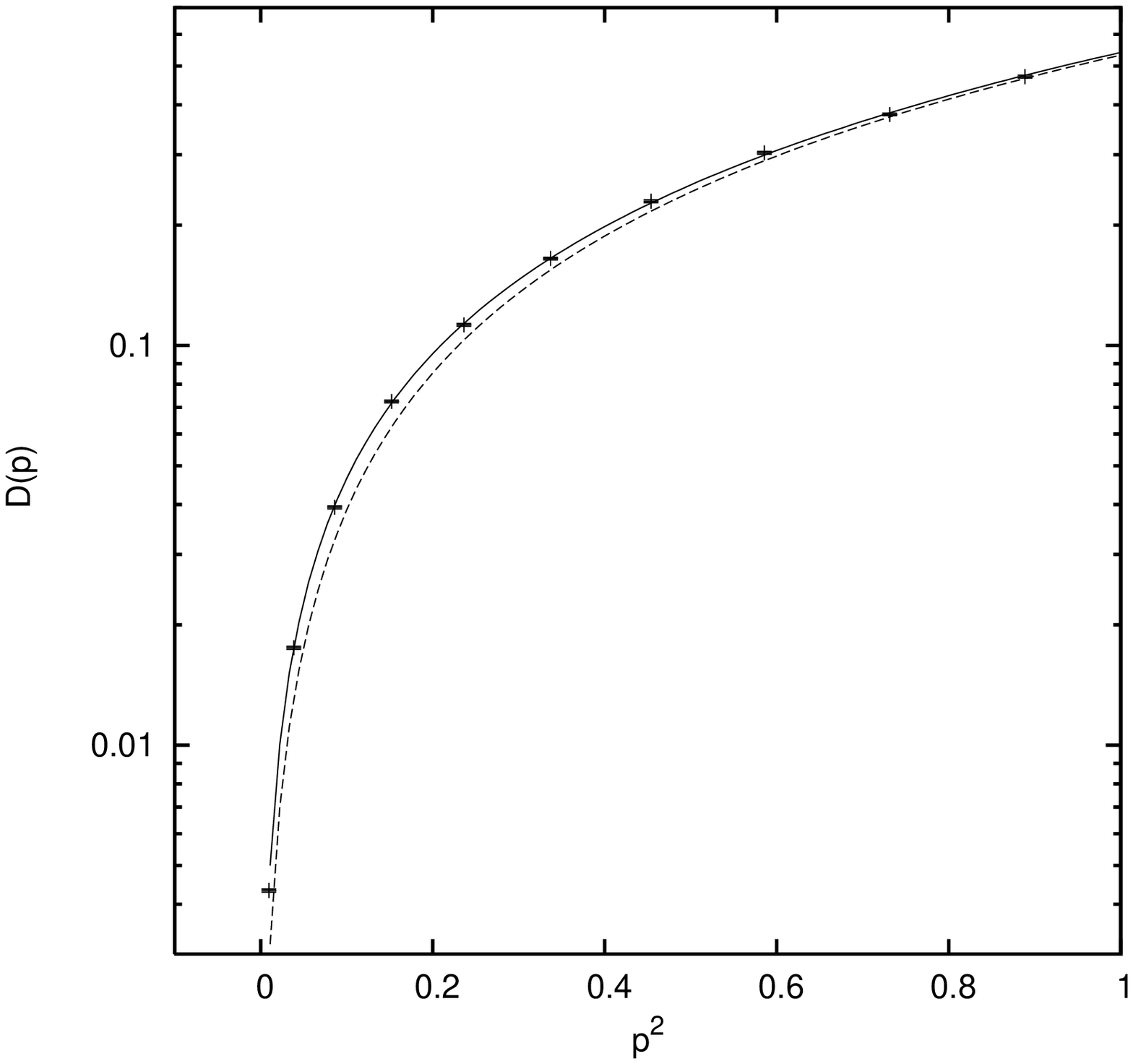}
\caption{\label{fig:4d-low}
  The gluon propagator $D(p)$ as a function of the (unimproved)
  lattice momenta $p^2$ for the lattice volume $V = 64^4$ for momenta of
  the type $(0,0,0,q)$ and $p^2 \leq 1$. We also present the fits obtained
  at small momenta $p^2 \leq 1.5$ (solid line) and at large momenta 
  $p^2 > 1.5$ (dashed line). See Tables \ref{tab:gluon-low-4d} and 
  \ref{tab:gluon-high-4d}.
  }
\end{figure}

\begin{figure}
\vspace{-44mm}
\includegraphics[scale=0.48]{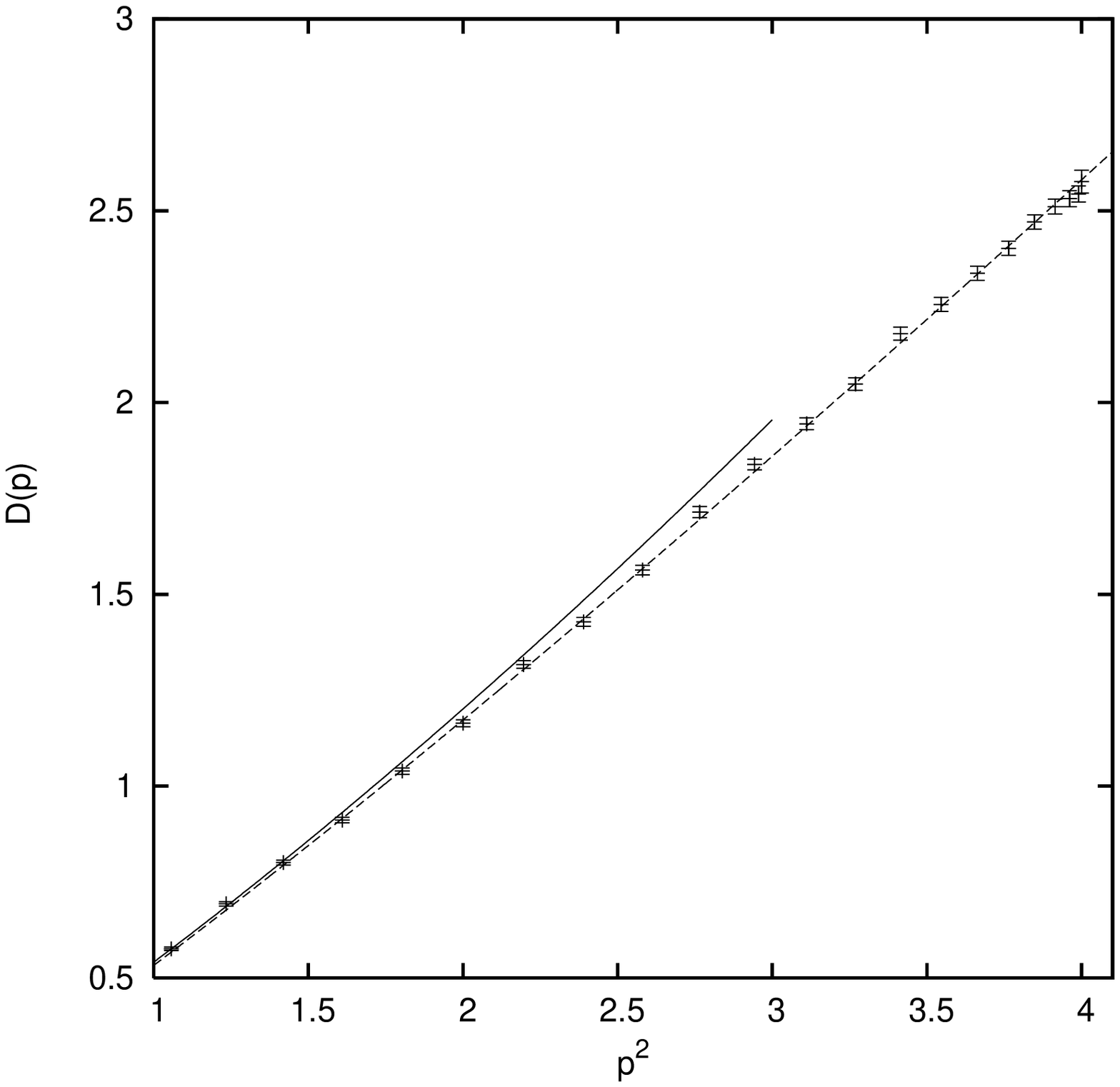}
\caption{\label{fig:4d-high}
  The gluon propagator $D(p)$ as a function of the (unimproved)
  lattice momenta $p^2$ for the lattice volume $V = 64^4$ for momenta of
  the type $(0,0,0,q)$ and $p^2 \geq 1$. We also present the fits obtained
  at small momenta $p^2 \leq 1.5$ (solid line) and at large momenta 
  $p^2 > 1.5$ (dashed line). See
  Tables \ref{tab:gluon-low-4d} and \ref{tab:gluon-high-4d}.
  }
\end{figure}

We did the same fitting analysis also in the 4d case for the gluon
propagator $D(p)$ at the lattice volume $V = 64^4$. As shown in Tables
\ref{tab:gluon-low-4d}, \ref{tab:gluon-high-4d} and \ref{tab:gluon-all-4d}
we find $D(0) \approx 0.45$ and $\kappa_Z \approx 0.9$ at small momenta
(using the fit with $b\neq 0$), while the fit for $p \geq 1.5$ (with $b=0$) 
gives $D(0)=0$ and $\kappa_Z \approx 0.56$.
(The fits corresponding to the range of small and large momenta are shown
together with the data in Figs.\ \ref{fig:4d-low} and \ref{fig:4d-high}.)
These results essentially do not change by cutting the data in Tables
\ref{tab:gluon-low-4d} and \ref{tab:gluon-high-4d} at $p^2 = 1$ or at $p^2 = 2$.
However, contrary to the 3d case, our data for $p^2 \geq 1.5$ can also be
described by the fitting function $b+c x^{2 \kappa_Z - 1}$ with $b \neq 0$.
Indeed, in this case, a good fit --- with $\chi^2/dof \approx 0.97$ and
$dof = 23$ --- is obtained with $b=0.4(1), c=0.14(9)$ and $\kappa_Z=0.67(6)$.
Finally, a fit using all the momenta and $b=0$ suggests a kind of average
value for $\kappa_Z$, i.e.\ $\kappa_Z \approx 0.7$. Thus, 
the fits suggest a value for $\kappa_Z$ close to the scaling solution result
$\kappa \approx 0.5953$ only if one ignores the data at small momenta and
forces the parameter $b$ to be zero.

\begin{figure}
\vspace{-50mm}
\includegraphics[scale=0.48]{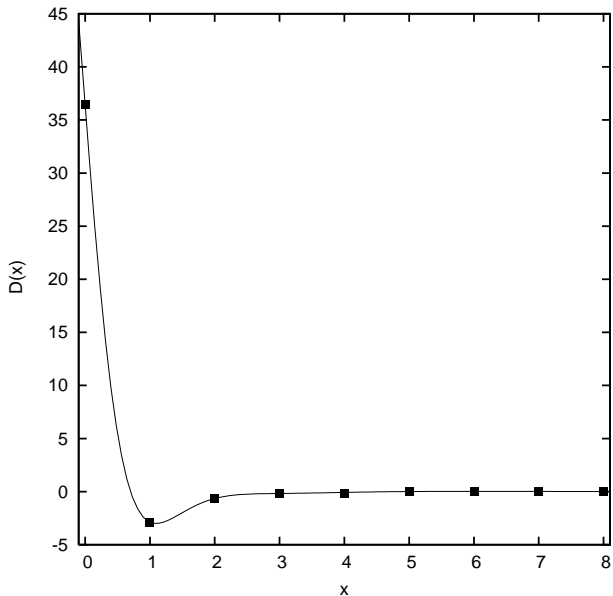}
\caption{\label{fig:4d-t}
  The gluon propagator $D(x)$ as a function of the space-time separation
  $x$ for the lattice volume $V = 64^4$, obtained by Fourier transforming
  data corresponding to momenta of the type $(0,0,0,q)$. For clarity, only the interval
  $x \in [0,8]$ is represented here. We also present the fit obtained
  using the fitting function $f(x)$ described in the text with
  $c_1 = D(x=0) = 36.4604, b_1 = 0, b_2 = \pi/2, \lambda_1 = 2.15(2),
  c_2 = 1.7(1)$ and $\lambda_2 = 0.65(3)$.
  }
\end{figure}

As in the 3d case, the effects due to violation of rotational symmetry
are clearly small. Also, violation of reflection positivity
is observed already at $x = 1$ and the gluon propagator in position space
$D(x)$ is well described by a sum of two Stingl-like propagators $f(x) = f_1(x) +
f_2(x)$ with $f_i(x) = c_i \, \cos{(b_i + \lambda_i x)} \, e^{-\lambda_i x}$
(see Fig.\ \ref{fig:4d-t}). Note that the values of the masses $\lambda_i$ are not
very different from the 3d case. Finally, as explained in Section \ref{sec:3dgluon}
above, one can fix the lattice spacing by comparing these data to results obtained
in the scaling region. In particular, considering our data at $\beta = 2.2$ for
$V = 128^4$, we find that $D(x) \approx 0$ for $x \gtapprox 2 fm$. Then, from Fig.\
\ref{fig:4d-t}, we find $a \approx 1 fm$ as in the 3d case.

\begin{table}
\vskip -2mm
\caption{Parameter fit for the gluon propagator $D(p)$ as a function of
         the (unimproved) lattice momentum $p^2$ using the fitting function
         $b+c (p^2)^{2 \kappa_Z-1}$ and data points in the range $p^2 \in
         [0,1.5]$ for the lattice volume $64^4$. We do a separate fit for
         each of the four types of momenta $(0,0,0,q)$, $(0,0,q,q)$, $(0,q,q,q)$
         and $(q,q,q,q)$, here indicated (respectively) as 1,2, 3 and 4.
         For each fit we also report the number of degrees of
         freedom ($dof$) of the fit and the value of $\chi^2/dof$.
\label{tab:gluon-low-4d}}
\vskip 2mm
\begin{tabular}{cccccc}
\hline
 momenta &   $b$    &   $c$    &   $\kappa_Z$    & $dof$ & $\chi^2 / dof$ \\
 \hline
 1       & 0.446(4) & 0.095(5) &  0.85(3)  &  11  &  1.33  \\
 2       & 0.460(5) & 0.095(6) &  0.83(4)  &  6  &  0.85  \\
 3       & 0.41(4)  & 0.14(4)  &  0.67(8)  &  4  &  1.28  \\
 4       & 0.45(2)  & 0.08(2)  &  0.9(2)   &  3  &  1.97  \\
 \hline
\end{tabular}
\vskip 1mm
\end{table}

\begin{table}
\vskip -2mm
\caption{Parameter fit for the gluon propagator $D(p)$ as a function of
         the (unimproved) lattice momentum $p^2$ using the fitting function
         $c (p^2)^{2 \kappa_Z-1}$ and data points in the range $p^2 \geq 1.5$
         for the lattice volume $64^4$. We do a separate fit for
         each of the four types of momenta $(0,0,0,q)$, $(0,0,q,q)$, $(0,q,q,q)$
         and $(q,q,q,q)$, here indicated (respectively) as 1,2, 3 and 4.
         For each fit we also report the number of degrees of
         freedom ($dof$) of the fit and the value of $\chi^2/dof$.
\label{tab:gluon-high-4d}}
\vskip 2mm
\begin{tabular}{cccccc}
\hline
 momenta &   $c$    &   $\kappa_Z$    & $dof$ & $\chi^2 / dof$ \\
 \hline
 1       & 0.533(3) &  0.570(3) &  17  &  0.76  \\
 2       & 0.558(4) &  0.562(2) &  21  &  0.86  \\
 3       & 0.561(6) &  0.561(3) &  23  &  0.99  \\
 4       & 0.531(6) &  0.566(2) &  24  &  0.97  \\
 \hline
\end{tabular}
\vskip 1mm
\end{table}

\begin{table}
\vskip -2mm
\caption{Parameter fit for the gluon propagator $D(p)$ as a function of
         the (unimproved) lattice momentum $p^2$ using the fitting function
         $b+c (p^2)^{2 \kappa_Z-1}$ and all data points for the lattice
         volume $64^4$. We do a separate fit for
         each of the four types of momenta $(0,0,0,q)$, $(0,0,q,q)$, $(0,q,q,q)$
         and $(q,q,q,q)$, here indicated (respectively) as 1,2, 3 and 4.
         For each fit we also report the number of degrees of
         freedom ($dof$) of the fit and the value of $\chi^2/dof$.
\label{tab:gluon-all-4d}}
\vskip 2mm
\begin{tabular}{ccccccc}
\hline
 momenta &  $b$  &   $c$     &    $\kappa_Z$    & $dof$ & $\chi^2 / dof$ \\
 \hline
 1       &  0.436(4) &  0.103(5) &  0.76(1)  &  30  &  1.60  \\
 2       &  0.430(9) &  0.13(1)  &  0.70(1)  &  29  &  2.20  \\
 3       &  0.39(2)  &  0.17(2)  &  0.65(1)  &  29  &  1.38  \\
 4       &  0.41(1)  &  0.13(2)  &  0.68(1)  &  29  &  1.19  \\

 \hline
\end{tabular}
\vskip 1mm
\end{table}

\subsubsection{3d ghost propagator} 

We now consider the data obtained for the ghost propagator in the 3d case.
[Note that in this case we did not evaluate the propagator for the lattice
volume $V=30^3$.]
As said above, we tried a fit using the function $c (p^2)^{-\kappa_G-1}$.
Results are reported in Table \ref{tab:ghost-all}. One clearly sees that
the infrared exponent $\kappa_G$ decreases as the lattice volume increases,
in agreement with \cite{Cucchieri:1997dx,Cucchieri:2007md}. In this case
the effects due to the violation of rotational symmetry are more evident.
Indeed, the exponent $\kappa_G$ is systematically smaller for momenta
along the axes. This is probably due to the fact that along the axes one
can consider smaller momenta, for which the ghost propagator is less
enhanced.

\begin{figure}
\vspace{-45mm}
\includegraphics[scale=0.49]{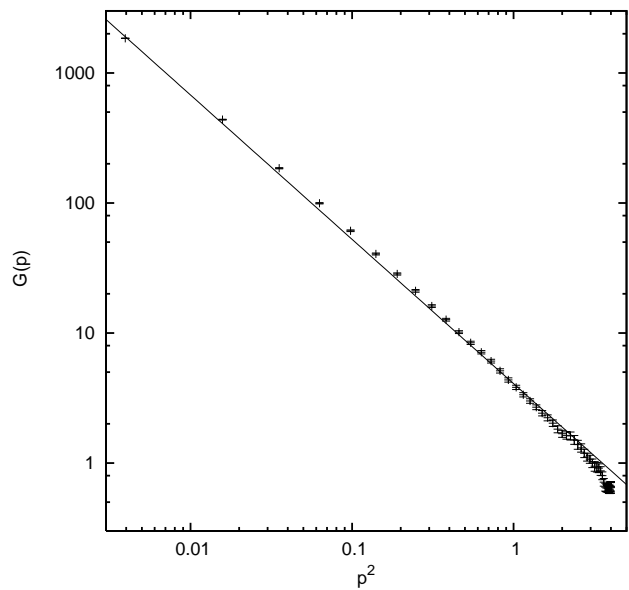}
\caption{\label{fig:3d-100-gh}
  The ghost propagator $G(p)$ as a function of the (unimproved)
  lattice momenta $p^2$ for the lattice volume $V = 100^3$ for momenta of
  the type $(0,0,q)$. We also present the corresponding fit reported in
  Table \ref{tab:ghost-all}.
  }
\end{figure}

\begin{figure}
\vspace{-46mm}
\includegraphics[scale=0.48]{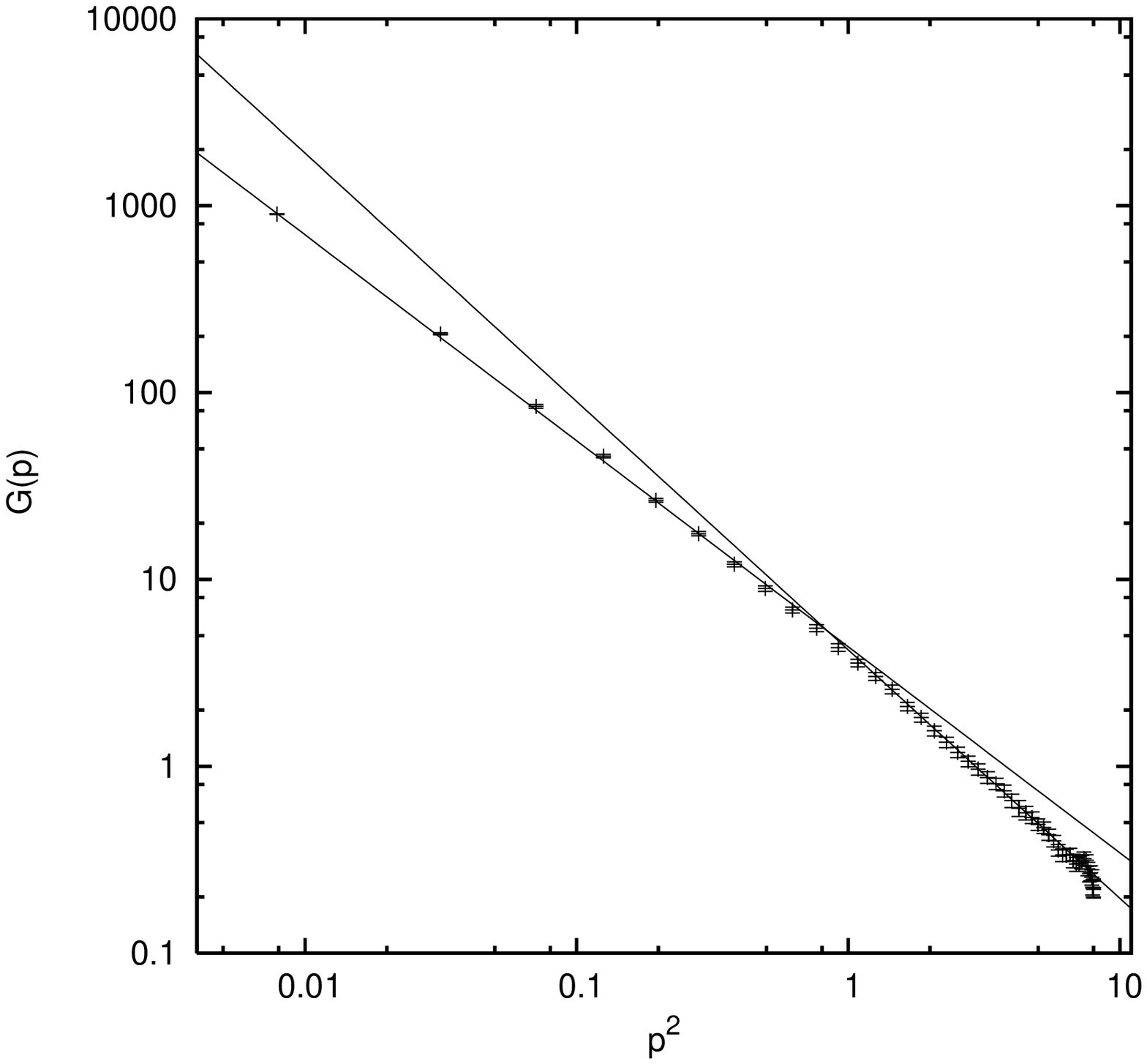}
\caption{\label{fig:3d-100-gh-bis}
  The ghost propagator $G(p)$ as a function of the (unimproved)
  lattice momenta $p^2$ for the lattice volume $V = 100^3$ for momenta of
  the type $(0,q,q)$. We also present the fits for the first and third
  sets of data points reported in Table \ref{tab:ghost-part}.
  }
\end{figure}

By looking at Fig.\ \ref{fig:3d-100-gh} it is clear that this fit, with
only one term, can be improved. In particular, considering the
results reported in \cite{Sternbeck:2008wg,Sternbeck:2008mv,Sternbeck:2008na},
one can try to see how the exponent $\kappa_G$ depends on the value of $p^2$.
To this end we have ordered the data points by the value of $p^2$ and
divided them in sets of ten data points. Then, we did a separate fit in
each interval, again using the fitting function $c (p^2)^{-\kappa_G-1}$.
Results for the lattice volume $V=100^3$ are reported in Table
\ref{tab:ghost-part}. One clearly sees that the exponent $\kappa_G$
increases with $p^2$ (see also Fig.\ \ref{fig:3d-100-gh-bis}) and it is
usually larger for the type of momenta $(0,q,q)$. The same is
observed for the other lattice volumes. This result is actually well-known.
Indeed, all lattice studies of the ghost propagator \cite{Cucchieri:1997dx,
Furui:2003jr,Gattnar:2004bf} have found that $G(p)$ is enhanced compared
to the tree-level propagator $1/p^2$ at intermediate momenta. (We will
comment again on this in the Conclusions.)

\begin{table}
\vskip -2mm
\caption{Parameter fit for the ghost propagator $G(p)$ as a function of
         the (unimproved) lattice momentum $p^2$ using the fitting function
         $c (p^2)^{-\kappa_G-1}$ and all data.
         We do a separate fit for each of the two types of momenta
         $(0,0,q)$, $(0,q,q)$, here indicated (respectively) as
         1 and 2. In each case we indicate the smallest nonzero
         momentum $p_{min}$. Finally, for each fit we also report the number
         of degrees of freedom ($dof$) of the fit and the value of $\chi^2/dof$.
\label{tab:ghost-all}}
\vskip 2mm
\begin{tabular}{cccccccc}
\hline
 $V$     & momenta &    $p_{min}$   &   $c$    &   $\kappa_G$    & $dof$ & $\chi^2 / dof$ \\
 \hline
 $10^3$  &    1    &    0.382       & 3.94(3)  & 0.260(8) &  3  &  0.39 \\
 $10^3$  &    2    &    0.764       & 3.9(1)   & 0.35(2)  &  3  &  1.75 \\
 \hline
 $20^3$  &    1    &    0.098       & 3.77(5)  & 0.209(7) &  8  &  4.15 \\
 $20^3$  &    2    &    0.196       & 3.57(6)  & 0.28(1)  &  8  &  3.06 \\
 \hline
 $40^3$  &    1    &    0.0246      & 3.64(8)  & 0.171(7) & 18 & 14.2 \\
 $40^3$  &    2    &    0.0492      & 3.38(7)  & 0.220(8) & 18 & 5.59 \\
 \hline
 $60^3$  &    1    &    0.0110      & 3.74(8)  & 0.144(5) & 28 & 29.2 \\
 $60^3$  &    2    &    0.0219      & 3.89(8)  & 0.192(8) & 28 & 16.7 \\
 \hline
 $80^3$  &    1    &    0.0062      & 3.84(9)  & 0.127(6) & 38 & 18.4 \\
 $80^3$  &    2    &    0.0123      & 3.50(7)  & 0.164(6) & 38 & 7.02 \\
 \hline
 $100^3$ &    1    &    0.0039      & 4.1(1)   & 0.110(5) & 48 & 19.4 \\
 $100^3$ &    2    &    0.0079      & 3.46(8)  & 0.154(5) & 48 & 9.91 \\
 \hline
\end{tabular}
\vskip 1mm
\end{table}

\begin{table}
\vskip -2mm
\caption{Parameter fit for the ghost propagator $G(p)$ as a function of
         the (unimproved) lattice momentum $p^2$ using the fitting function
         $c (p^2)^{-\kappa_G-1}$ and various sets of ten data points for the
         lattice volume $100^3$
         (we report in the table the smallest and the largest momenta
         considered for each set). We do a separate fit for each of the
         two types of momenta $(0,0,q)$, $(0,q,q)$, here indicated
         (respectively) as 1 and 2. For each fit we also report the
         value of $\chi^2/dof$ [the number of degrees of
         freedom ($dof$) is always 8]. We do not report the fit for
         the last set because the corresponding data points have large
         statistical fluctuations.
\label{tab:ghost-part}}
\vskip 2mm
\begin{tabular}{ccccccc}
\hline
 $p_{min}$ & $p_{max}$ & momenta &   $c$    &   $\kappa_G$    & $\chi^2 / dof$ \\
 \hline
 0.0039 & 0.38  &    1    & 5.0(2) & 0.070(6) &  11.2 \\
 0.0079 & 0.76  &    2    & 4.4(1) & 0.102(7) &  4.40 \\
 \hline
 0.46   & 1.38  &    1    & 4.00(2) & 0.21(1) &  0.17 \\
 0.92   & 2.76  &    2    & 3.98(5) & 0.29(2) &  0.19 \\
 \hline
 1.50   & 2.62  &    1    & 3.7(1)  & 0.08(5) &  0.19 \\
 3.01   & 5.24  &    2    & 4.2(1)  & 0.33(3) &  0.04 \\
 \hline
 2.74   & 3.62  &    1    & 6(1)    & 0.6(1)  &  0.31 \\
 5.47   & 7.24  &    2    & 3(1)    & 0.3(1)  &  0.28 \\
 \hline
\end{tabular}
\vskip 1mm
\end{table}


\subsubsection{4d ghost propagator} 

\begin{table}
\vskip -2mm
\caption{Parameter fit for the ghost propagator $G(p)$ as a function of
         the (unimproved) lattice momentum $p^2$ using the fitting function
         $c (p^2)^{-\kappa_G-1}$ and all data for the lattice volume $64^4$.
         We do a separate fit for each of the four types of momenta
         $(0,0,0,q)$, $(0,0,q,q)$, $(0,q,q,q)$, $(q,q,q,q)$, here indicated
         (respectively) as 1, 2, 3 and 4. In each case we indicate the smallest nonzero
         momentum $p_{min}$. Finally, for each fit we also report
         the value of $\chi^2/dof$ [the number of degrees of
         freedom ($dof$) is always 30].
\label{tab:ghost-4d-all}}
\vskip 2mm
\begin{tabular}{cccccc}
\hline
 momenta &    $p_{min}$   &   $c$    &   $\kappa_G$    & $\chi^2 / dof$ \\
 \hline
 1    &    0.0096      & 5.4(2)  &  0.075(7) &   250  \\
 2    &    0.0193      & 4.6(2)  &  0.128(9) &   238  \\
 3    &    0.0289      & 4.2(1)  &  0.16(1)  &   127  \\
 4    &    0.0385      & 4.0(1)  &  0.19(1)  &   55.3 \\
 \hline 
\end{tabular}
\vskip 1mm
\end{table}

\begin{table}
\vskip -2mm
\caption{Parameter fit for the ghost propagator $G(p)$ as a function of
         the (unimproved) lattice momentum $p^2$ using the fitting function
         $c (p^2)^{-\kappa_G-1}$ and various sets of eight data points for the
         lattice volume $64^4$ (we report in the table the smallest and the
         largest momenta considered for each set). We do a separate fit for
         each of the four types of momenta
         $(0,0,0,q)$, $(0,0,q,q)$, $(0,q,q,q)$, $(q,q,q,q)$, here indicated
         (respectively) as 1, 2, 3 and 4. For each fit we also report the
         value of $\chi^2/dof$ [the number of degrees of
         freedom ($dof$) is always 6].
\label{tab:ghost-part-4d}}
\vskip 2mm
\begin{tabular}{ccccccc}
\hline
 $p_{min}$ & $p_{max}$ & momenta &   $c$    &   $\kappa_G$    & $\chi^2 / dof$ \\
 \hline
 0.0096 & 0.59  &    1    & 6.4(2) & 0.036(7) & 101   \\
 0.0193 & 1.17  &    2    & 5.8(2) & 0.06(1)  & 114   \\
 0.0289 & 1.76  &    3    & 5.4(2) & 0.09(1)  & 65.4  \\
 0.0385 & 2.34  &    4    & 5.3(2) & 0.10(1)  & 18.6  \\
 \hline
 0.73 & 2  &        1     & 5.192(9) & 0.223(5) & 0.13 \\
 1.46 & 4  &        2     & 5.28(3)  & 0.300(6) & 0.11 \\
 2.19 & 6  &        3     & 5.64(5)  & 0.389(7) & 0.09 \\
 2.92 & 8  &        4     & 5.8(2)   & 0.39(2)  & 0.18 \\
 \hline
 2.20 & 3.41  &      1    & 5.78(8) & 0.38(1) &  0.06 \\
 4.39 & 6.83  &      2    & 5.94(9) & 0.403(9)&  0.03 \\
 6.59 & 10.2  &      3    & 5.8(2)  & 0.39(1) &  0.04 \\
 8.78 & 13.7  &      4    & 7.2(7)  & 0.51(4) &  0.10 \\
 \hline
 3.55  & 4  &       1     & 5.8(2)  & 0.39(2) &  0.01 \\
 7.09  & 8  &       2     & 5.8(4)  & 0.39(3) &  0.02 \\
 10.6  & 12 &       3     & 19(6)   & 0.9(1)  &  0.24 \\
 14.2  & 16 &       4     & 7(2)    & 0.5(1)  &  0.08 \\
 \hline
\end{tabular}
\vskip 1mm
\end{table}

\begin{figure}
\vspace{-46mm}
\includegraphics[scale=0.48]{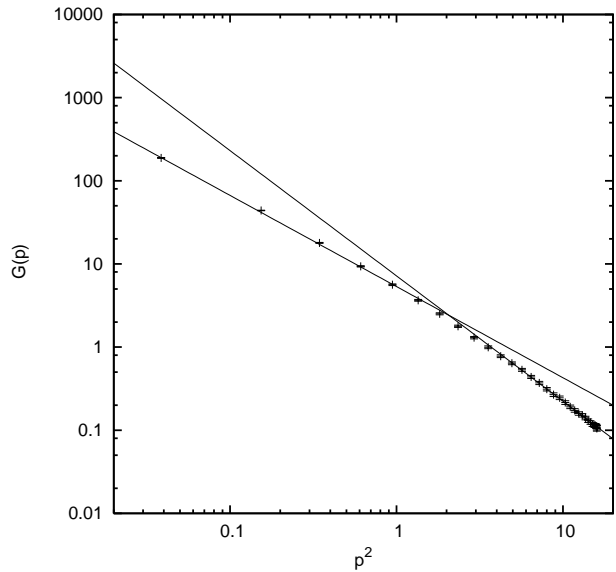}
\caption{\label{fig:4d-gh}
  The ghost propagator $G(p)$ as a function of the (unimproved)
  lattice momenta $p^2$ for the lattice volume $V = 64^4$ for momenta of
  the type $(q,q,q,q)$. We also present the fits for the first and third
  sets of data points reported in Table \ref{tab:ghost-part-4d}.
  }
\end{figure}

\begin{figure}
\includegraphics[scale=0.41,angle=-90]{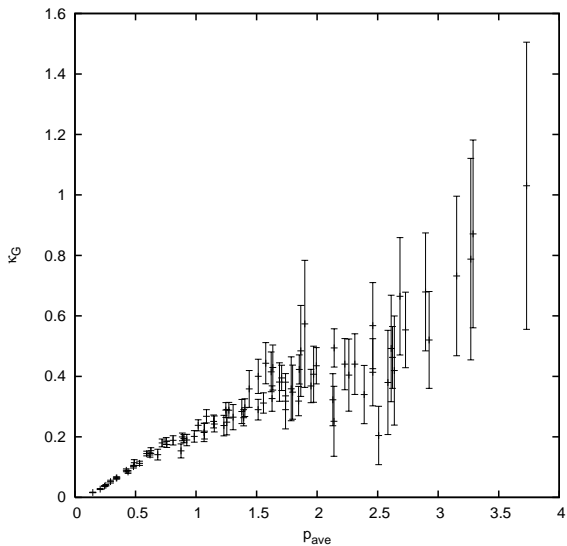}
\caption{\label{fig:4d-gh-effective-k}
  The effective ghost exponent $\kappa_G$, evaluated using
  $-1+0.5\log{[G(p_1)/G(p_2)]}/log{(p_2/p_1)}$,
  as a function of the average momentum $p_{ave} = (p_1 + p_2)/2$
  for the lattice volume $V = 64^4$. Here,
  $G(p_1)$ and $G(p_2)$ are the values of the ghost propagator
  at two nearby momenta $p_1$ and $p_2$.
  }
\end{figure}

\begin{figure}
\includegraphics[scale=0.41,angle=-90]{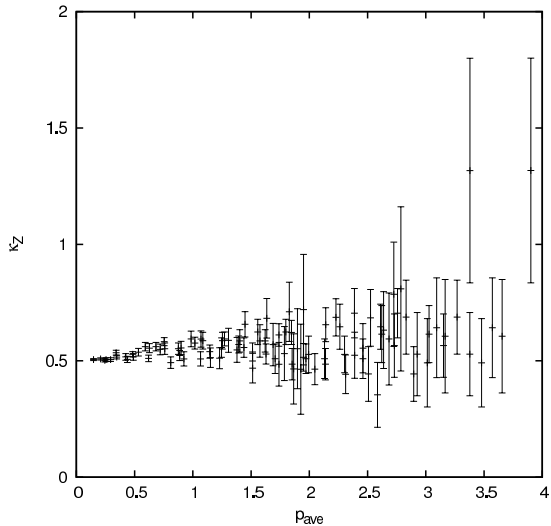}
\caption{\label{fig:4d-gl-effective-k}
  The effective gluon exponent $\kappa_Z$,
  evaluated using $0.25 \, \{ 2 + \log{[D(p_1)/D(p_2)]} /
  log{(p_1/p_2)} \}$,
  as a function of the average momentum $p_{ave} = (p_1 + p_2)/2$
  for the lattice volume $V = 64^4$. Here,
  $D(p_1)$ and $D(p_2)$ are the values of the gluon propagator
  at two nearby momenta $p_1$ and $p_2$.
  }
\end{figure}

Finally, data for the ghost propagator $G(p)$ in the 4d case show
again (see Table \ref{tab:ghost-4d-all}) that the exponent $\kappa_G$
depends on the type of momenta and systematically increases when
one considers momenta closer to the diagonal than to the axes.
Also, the exponent $\kappa_G$ increases with $p^2$ (see Table
\ref{tab:ghost-part-4d} and Fig.\ \ref{fig:4d-gh}), going from a very small value ---
close to zero --- at small $p^2$ up to almost 1 for the largest
momenta and for momenta along the diagonal. This can be seen also
in Fig.\ \ref{fig:4d-gh-effective-k}, where an effective exponent
$\kappa_G$ has been evaluated using the relation $-1+0.5\log{[G(p_1)/G(p_2)]}/log{(p_2/p_1)}$,
where $G(p_1)$ and $G(p_2)$ are the values of the ghost propagator
at two nearby momenta $p_1$ and $p_2$. The plot is done as a function
of the average momentum $p_{ave} = (p_1 + p_2)/2$. The errors have been evaluated
using the so-called bootstrap method with 2000 samples. For clarity, we
show only results with a relative error smaller than 50\% and with
a positive value for $\kappa_G$ (this criterion
selects about 70\% of the data). One clearly sees that the effective
exponent $\kappa_G$ is monotonically increasing. The situation is very different
in the gluon sector, where indeed a scaling solution can be used to fit
the data. As one can see in Fig.\ \ref{fig:4d-gl-effective-k}, the
effective exponent $\kappa_Z = 0.25 \, \{ 2 + \log{[D(p_1)/D(p_2)]} / 
log{(p_1/p_2)} \}$ is essentially constant in this case. Moreover, at
small momenta one finds $\kappa_Z = 0.5$, in agreement with the massive
solution, and at large momenta the effective exponent is still very
close to this value. Note that a constant shift of 0.5 is built into
the definition of the gluon exponent $\kappa_Z$. Thus, numerical results
in the gluon sector should be rather quoted as $\kappa_Z-0.5$, in order to
convey a clear indication of the precision of the results.

\subsubsection{A massive fit for the ghost propagator} 
\label{sec:massive}

The analysis above has shown that the scaling solution does
not describe the ghost sector. Indeed, one cannot find a reasonably
large range of momenta where the data can be fitted by
a power-law with a given value of the IR exponent $\kappa_G$.

\begin{figure}
\vspace{-43mm}
\includegraphics[scale=0.48]{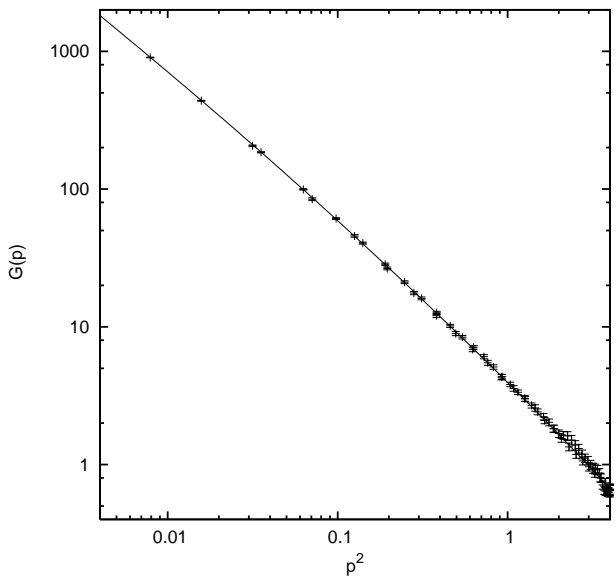}
\caption{\label{fig:all-3d-ghost}
  The ghost propagator $G(p)$ as a function of the (unimproved)
  lattice momenta $p^2$ for the lattice volume $V = 100^3$ and
  momenta $p^2 \leq 4$. We also show a fit using the fitting
  function $f(x) = \left[ a - b \log{(p^2 + c^2)} \right] / p^2$, discussed in
  the text.
  }
\end{figure}

\begin{figure}
\vspace{-43mm}
\includegraphics[scale=0.48]{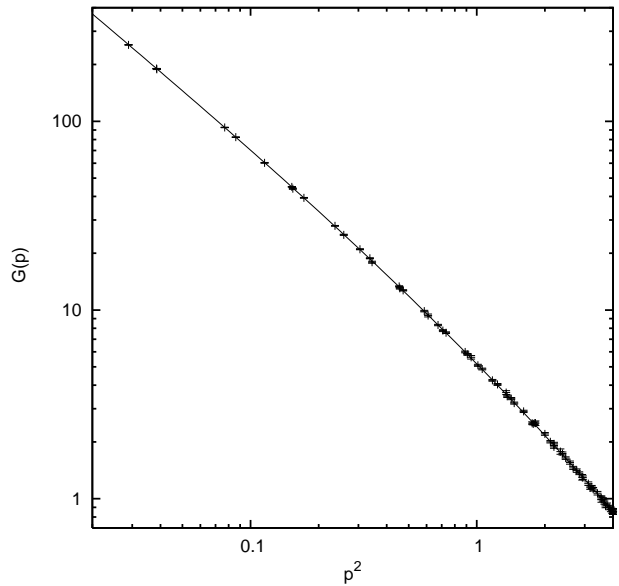}
\caption{\label{fig:all-4d-ghost}
  The ghost propagator $G(p)$ as a function of the (unimproved)
  lattice momenta $p^2$ for the lattice volume $V = 64^4$ and
  momenta $p^2 \leq 4$. We also show a fit using the fitting
  function $f(x) = \left[ a - b \log{(p^2 + c^2)} \right] / p^2$, discussed in
  the text.
  }
\end{figure}

\begin{figure}
\vspace{-43mm}
\includegraphics[scale=0.48]{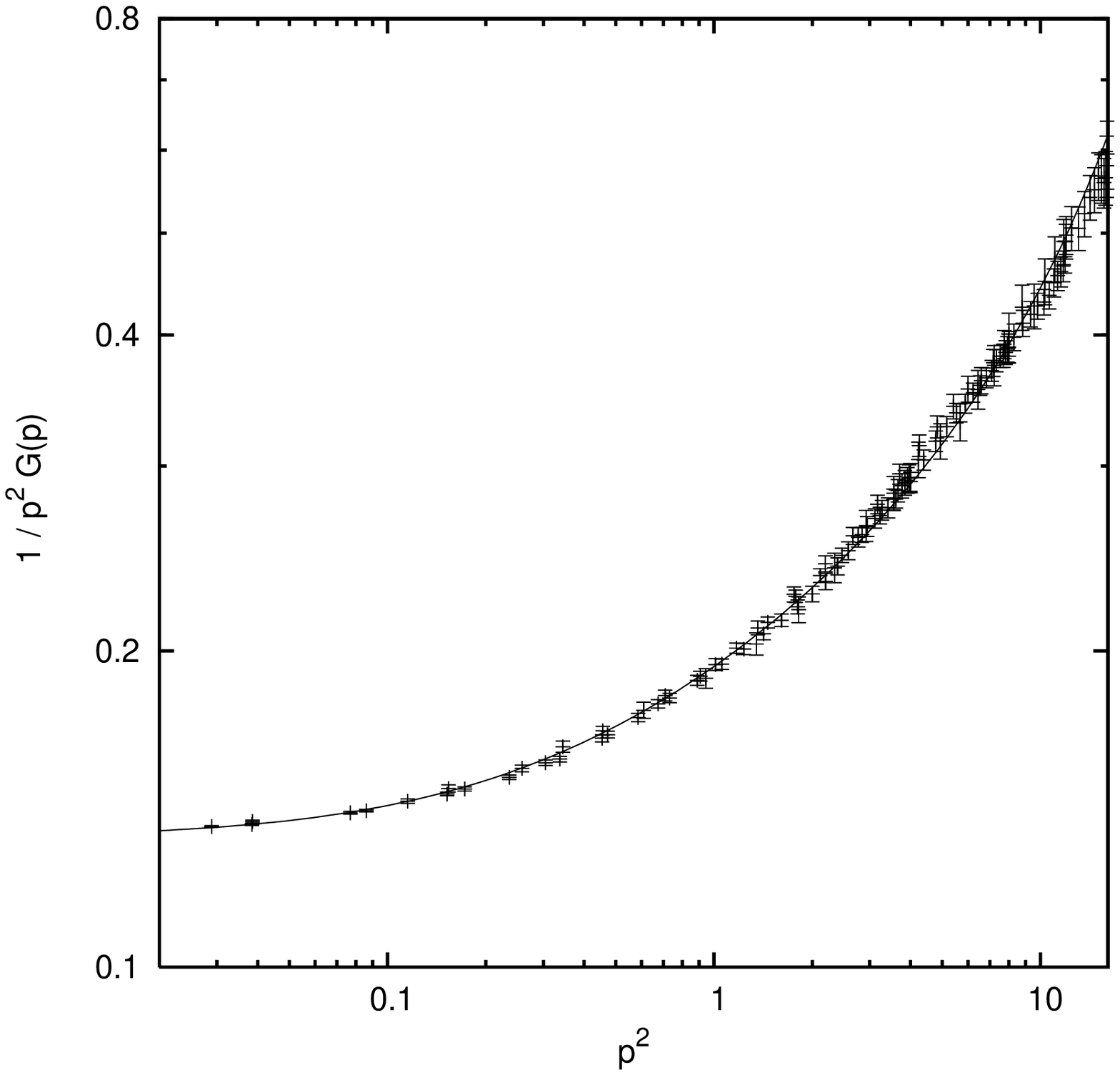}
\caption{\label{fig:all-4d-ghost-Z22-log}
  The inverse ghost dressing function $1 / [ p^2 G(p) ]$ as a function of the (unimproved)
  lattice momenta $p^2$ for the lattice volume $V = 64^4$.
  We also show a fit using the fitting function
  $1 / \left[ a - b \log{(p^2 + c^2)} \right]$, discussed in
  the text.
  }
\end{figure}

On the other hand, the ghost-propagator data are well described by the
fitting function $f(x) = \left[ a - b \log{(p^2 + c^2)} \right] / p^2$, recently
proposed in \cite{Aguilar:2008xm}, which gives a free ghost propagator
in the infrared limit. The parameter $c$ can be interpreted as a gluon mass
\cite{Aguilar:2008xm}. Also note that this function corresponds to the
small-momentum limit of the fitting function used in Ref.\
\cite{Cucchieri:2008fc} to fit the ghost data in 3d and in 4d
for values of $\beta$ in the scaling region. Using the fitting function
$f(x)$ above we obtain, in the 3d case, the parameters $a = 3.96(2)$,
$b=0.92(2)$ and $c=0.155(6)$ with $ \chi^2/dof = 0.73$ using the data
$p^2 \leq 4$ (see Fig.\ \ref{fig:all-3d-ghost}). Using all the data one finds
$a=3.95(2)$, $b=0.94(1)$ and $c=0.159(6)$ with $ \chi^2/dof = 0.72 $.
A similar fit in the 4d case gives $a = 5.51(2)$,
$b=1.45(1)$ and $c=0.499(7)$ with $ \chi^2/dof = 0.65$ using the data $p^2 \leq 4$
(see Fig.\ \ref{fig:all-4d-ghost}). Using all the data one finds $a=5.44(1)$,
$b=1.372(8)$ and $c=0.466(5)$ with $ \chi^2/dof = 0.93 $ (see
Fig.\ \ref{fig:all-4d-ghost-Z22-log}).

\begin{figure}
\vspace{-43mm}
\includegraphics[scale=0.48]{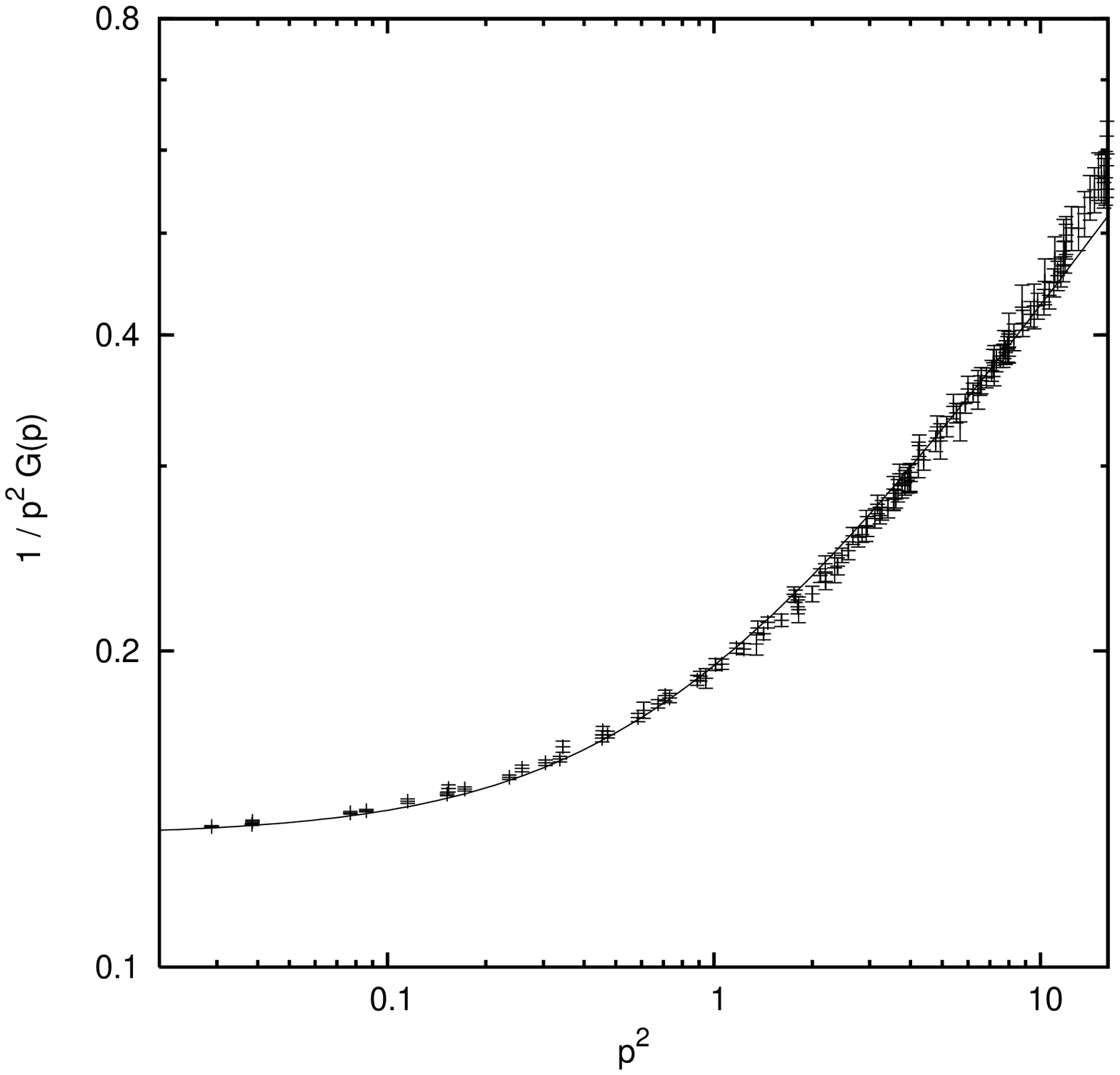}
\caption{\label{fig:all-4d-ghost-Z11-log}
  The inverse ghost dressing function $1 / [ p^2 G(p) ]$ as a function of the (unimproved)
  lattice momenta $p^2$ for the lattice volume $V = 64^4$. 
  We also show a fit using the fitting function
  $g(x) = (p^2 + l)^{\kappa_G} / e$, discussed in
  the text. 
  }
\end{figure}

\begin{figure}
\vspace{-43mm}
\includegraphics[scale=0.48]{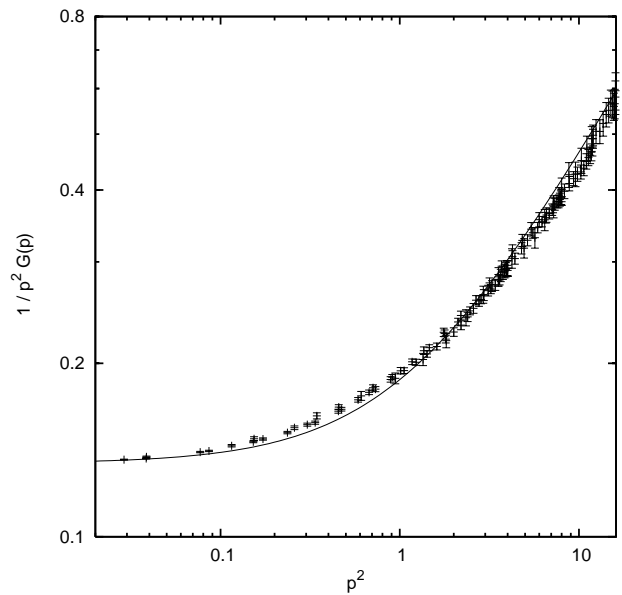}
\caption{\label{fig:all-4d-ghost-Z33-log}
  The inverse ghost dressing function $1 / [ p^2 G(p) ]$ as a function of the (unimproved)
  lattice momenta $p^2$ for the lattice volume $V = 64^4$.
  We also show a fit using the fitting function
  $g(x) = (p^2 + l)^{0.562}  / e$, discussed in
  the text.
  }
\end{figure}

For a comparison with Ref.\ \cite{Sternbeck:2008mv} ---
see Eq.\ (19c) and Fig.\ 9 of that reference --- we have also
tried a global fit using for the ghost dressing function
the Ansatz $g(x) = e / (p^2 + l)^{\kappa_G}$. The resulting fit gives
$e = 3.88(3), l=0.098(8), \kappa_G = 0.271(7)$ with $ \chi^2/dof = 1.63$ in 3d and
$e = 6.56(7), l=0.73(2), \kappa_G = 0.434(6)$ with $ \chi^2/dof = 1.79$
in the 4d case (see Fig.\ \ref{fig:all-4d-ghost-Z11-log}).
Thus, the value of $ \chi^2/dof $ is clearly worse when compared to
the logarithmic behavior considered above.
In particular, from Fig.\ \ref{fig:all-4d-ghost-Z11-log}
one can see that the power-law behavior cannot describe well
the ``curvature'' of the data, underestimating the data
at small and at large momenta and overshooting the numerical
results at intermediate momenta. This is of course not a surprise
since we have clearly shown in the previous subsection that a
single power-law does not describe the ghost data, unless one
selects a very small interval of momenta. In the 4d case, this result
does not improve if one forces the exponent $\kappa_G$ to be equal to
0.562, as done in Fig.\ 9 of \cite{Sternbeck:2008mv}. Indeed, in this case we
find $e = 8.37(6)$ and $l = 1.23(2)$ with $ \chi^2/dof = 5.7$
(see Fig.\ \ref{fig:all-4d-ghost-Z33-log}).

One should also observe that the function $g(x)$ considered above and
in Ref.\ \cite{Sternbeck:2008mv} is not
a truly scaling solution $h(x) = s /(p^2)^{\kappa_G}$. Indeed, while the
latter is characterized by a constant value for the exponent 
$\kappa_G = - \partial \log [h(x)] / \partial \log [p^2] =
- [p^2/h(x)] \partial h(x) / \partial p^2$, for $g(x)$ one has
an effective exponent $- [p^2/g(x)] \partial g(x) / \partial p^2 =
\kappa_G \, p^2 / (p^2 + l)$. This effective exponent is monotonically
increasing with $p^2$, becoming constant and equal to $\kappa_G$ only for
very large momenta. Thus, $g(x)$ is trying to describe the lattice data,
characterized by an exponent $\kappa_G$ increasing with the momentum,
while ``suggesting'' a possible scaling behavior, i.e.\ a function
with a constant exponent. In fact, as we have shown
above, the data are much better described by the massive solution $f(x)$,
suggested by a recent analytic study \cite{Aguilar:2008xm}. On the other hand, a function
such as $g(x)$ is very poorly justified from the theoretical point
of view. In particular, such a fitting function for the ghost dressing
function implies a fit $e / [p^2 (p^2 + l)^{\kappa_G}]$ for the ghost
propagator. We do not see any theoretical reason that the mass scale $\sqrt{l}$
should affect only ``part'' of the power-law behavior of the propagator.
Here we have also tried a fit to the ghost-propagator data using
$e / [(p^2 + l)^{1+\kappa_G}]$:
the result is very poor, with $ \chi^2/dof = 95.6$ [and $\kappa_G = 0.182(6)$].

Finally, that a simple power-law is not capable of describing the lattice data
should also be clear looking at Fig.\ 5 of Ref.\ \cite{Sternbeck:2008mv},
where it is evident that the exponent $\kappa_G$ depends on the momentum $x$
considered, and from Fig.\ 9 of the same reference, where one sees
that the fit systematically underestimates the data at intermediate
momenta. Actually, at the end of Sec.\ III of Ref.\ \cite{Sternbeck:2008mv}
the authors clearly say that a true scaling solution, i.e.\ their
Eq.\ (19a), gives a very poor description of the ghost data and that their
preferred fitting function is $g(x)$, which only reminds one of a possible scaling
behavior. This is probably the reason that induced the authors of
\cite{Sternbeck:2008mv} to conclude in favor of a scaling solution at large momenta
in the ghost sector. As already stressed above, we do not agree with this conclusion.

\section{Conclusions}
\label{sec:conclusions}

We have studied numerically the infrared behavior of $SU(2)$ Landau-gauge gluon 
and ghost propagators at lattice parameter $\beta = 0$, considering 3d
lattices of volumes up to $100^3$ and 4d lattices of volume $64^4$.
By carrying out a careful fit analysis of the proposed behavior according to the
scaling or to the massive solutions of DSE, we find that our data strongly
support the massive solution, i.e.\ a finite gluon propagator and an essentially
free ghost propagator in the infrared limit $p \to 0$. Moreover, the gluon
propagator $D(x)$ as a function of the space-time separation $x$ violates
reflection positivity and it is well-described by a sum of two Stingl-like
propagators. These results are in qualitative agreement with data obtained at
finite $\beta$ in the scaling region.

We should stress that, in agreement with Refs.\ \cite{Sternbeck:2008wg,
Sternbeck:2008mv,Sternbeck:2008na}, a scaling solution appears in the gluon sector
if one neglects the data at small momenta. As explain in Section
\ref{sec:recent0} above, we do not see any reason for excluding those
data from the analysis. In particular, discretization effects at small
momenta are under control and the large effects observed in Ref.\
\cite{Sternbeck:2008mv} are probably only due to the bad scaling properties
of the modified Landau gauge.
Moreover, the value of $\kappa$ clearly depends on the way the
fits are done. In particular, a value close to the preferred value of the
scaling solution, i.e.\ $\, 0.2(d-1)$ in $d$ dimensions, is obtained only with very
specific and ad-hoc fits. In any case, we believe that the scaling solution is clearly
excluded by the ghost sector and we definetively do not agree on this point
with the analysis and the conclusions presented in \cite{Sternbeck:2008wg,
Sternbeck:2008mv,Sternbeck:2008na}. Indeed, in this case, the IR exponent $\kappa$
depends on the interval considered, increasing essentially monotonically as the
momentum increases, i.e.\ it is impossible to find a decent ``window'' giving
a constant value for $\kappa$. On the other hand, we have shown that the data for the
ghost propagator are very well described by a simple function, recently suggested
by an analytic study presented in \cite{Aguilar:2008xm}. This function
clearly supports the so-called massive solution.

As for the ongoing discussion about massive solution versus conformal scaling,
we remark that the lattice results may be summarized as follows.
\begin{itemize}
\item In 2d Landau gauge one sees conformal scaling \cite{Maas:2007uv,
      Cucchieri:2007rg,Cucchieri:2008fc},
\item In 3d and 4d Landau gauge one finds the massive solution
      \cite{Cucchieri:2007md,Bogolubsky:2007ud,Sternbeck:2007ug,
      Cucchieri:2008fc},
\item In 4d Coulomb gauge, the transverse gluon propagator is well
      described by a Gribov formula, going to zero at zero
      momentum \cite{Cucchieri:2000gu,Cucchieri:2000kw,Burgio:2008jr}
\item In the so-called $\lambda$ gauges, which interpolate between the
      Landau gauge ($\lambda = 1$) and the Coulomb gauge ($\lambda = 0$),
      one clearly sees \cite{Cucchieri:2007uj} that the behavior of the
      transverse gluon propagator gets modified when $\lambda$ goes from
      1 to 0, becoming closer and closer to the behavior obtained in
      Coulomb gauge as $\lambda$ becomes smaller and smaller.
\end{itemize}
The simulations cited above are essentially all done in the same way, i.e.\
in most of the cases by ignoring effects due to Gribov copies, by using a standard
discretization for the lattice action, for the gluon field and for the gauge-fixing
condition and by using one of the standard gauge-fixing algorithms. Recently, to
rescue the conformal solution in Landau gauge, several authors have
explained the lattice Landau data by evoking supposedly (very) large effects 
due to
Gribov copies, discretization effects and bias related to the use of the
usual gauge-fixing algorithms. Of course, one has to verify that all possible
sources of systematic effects are indeed under control in a numerical
simulation. On the other hand, it seems unlikely to us that these effects
would show up in some cases of the simulations described above and not in 
others. For example, why would the 2d Landau-gauge case be conformal and not
the 3d and 4d cases
if the same code is used in these three cases? Why should Gribov-copy
effects be so important in 3d and 4d Landau gauge and not in 2d Landau
gauge and, even more strikingly, in 4d Coulomb gauge, where one would expect stronger
effects since the transversality condition is imposed separately on each time slice?
In our view, present lattice data are simply telling us that the
infrared behavior of gluon and ghost propagators in Landau, Coulomb and $\lambda$
gauge depends on the gauge-fixing condition and on the dimensionality of the
system (as, for example, the critical behavior of statistical mechanical
systems). We believe that the present challenge is to understand why this is
the case and that the bounds introduced by us in Ref.\ 
\cite{Cucchieri:2007rg} and their interpretation in terms of magnetization
and susceptibility of the gluon field could be a key ingredient in a simple
explanation of present lattice results in Landau gauge.

Finally a remark about color confinement. As said in the Conclusion of
Ref.\ \cite{Cucchieri:2008fc}, we point out that the behavior of gluon and
ghost propagators at very small momenta is probably not so important
for the explanation of confinement. Indeed, why should the behavior of a Green
function at a few $MeV$, i.e.\ for a space separation of about $50 fm$,
be relevant for hadron physics, since the typical hadronic scale
is of order of $1 fm$? Let us recall that in a recent paper \cite{Yamamoto:2008am}
it has been shown that the appearance of a linearly-rising potential
is related (in Landau and in Coulomb gauge) to the momentum-space
gluon configuration $A(p)$ for $p \ltapprox 1 \, GeV$. In this region one can
indeed observe strong non-perturbative effects in the gluon and in the
ghost propagators in Landau gauge: the gluon propagator violates reflection
positivity and the ghost propagator is enhanced when compared to the
tree-level behavior. Thus, some important predictions of the Gribov-Zwanziger
scenario are still verified and, at the same time, one can try to relate the
massive solution to the requirement of color confinement as in
Refs.\ \cite{Chaichian:2006bn,Cornwall:1979hz,Braun:2007bx}.


\section{Acknowledgements}

We thank the organizers of the workshop
on Non perturbative Aspects of Field Theories (Morelia '09), where
this work was presented, for providing a very stimulating atmosphere
for discussions, in such a beautiful setting.

We acknowledge partial support from the Brazilian Funding
Agencies FAPESP and CNPq. The work of T.M.\ is supported also
by a fellowship from the Alexander von Humboldt Foundation.
Most of the simulations reported here have been done on the
IBM supercomputer at S\~ao Paulo University (FAPESP grant \# 04/08928-3).


\end{document}